\documentclass[11pt,a4paper]{article}
\pdfoutput=1
\usepackage{jheppub}
\usepackage{amssymb}
\usepackage{amsfonts}
\usepackage{graphicx}
\usepackage{epstopdf}
\usepackage{dcolumn}
\usepackage{amsmath}
\usepackage{latexsym,bm}
\usepackage{amsthm}
\usepackage{slashed}
\usepackage{float}
\usepackage{color}

\def \be {\begin{equation}}
\def \ee {\end{equation}}
\def \bsp {\begin{split}}
\def \esp {\end{split}}
\def \bea {\begin{eqnarray}}
\def \eea {\end{eqnarray}}

\def\mc{\mathcal}

\def\ge{{\mathfrak{e}}}
\def\gso{{\mathfrak{so}}}
\def\gsu{{\mathfrak{su}}}
\def\gsp{{\mathfrak{sp}}}
\def\gf{{\mathfrak{f}}}
\def\gg{{\mathfrak{g}}}

\def\P{\mathbb{P}}

\def\Z{\mathbb{Z}}
\def\F{\mathbb{F}}

\title{A Monte Carlo exploration of
threefold base geometries for 4d F-theory vacua}

\author[a]{Washington Taylor,}
\author[a]{Yi-Nan Wang}

\affiliation[a]{Center for Theoretical Physics,\\Department of Physics\\Massachusetts Institute of Technology\\77 Massachusetts Avenue\\Cambridge, MA 02139, USA}

\emailAdd{wati@mit.edu}
\emailAdd{wangyn@mit.edu}

\preprint{\today \hspace*{0.1in} MIT-CTP-4697}

\abstract{
We use Monte Carlo methods to explore the set of toric threefold bases
that support elliptic Calabi-Yau fourfolds for F-theory
compactifications to four dimensions, and study the distribution of
geometrically non-Higgsable gauge groups, matter, and quiver
structure.  We estimate the number of distinct threefold bases in the
connected set studied to be $\sim { 10^{48}}$.  
The distribution of bases peaks around $h^{1, 1}\sim 82$.
All bases encountered
after ``thermalization'' have some geometric non-Higgsable structure.
We find that the number of non-Higgsable gauge group factors grows
roughly linearly in $h^{1,1}$ of the threefold base.  Typical bases
have $\sim 6$ isolated gauge factors as well as several larger
connected clusters of gauge factors with jointly charged matter.
Approximately 76\% of the bases sampled contain connected two-factor
gauge group products of the form SU(3)$\times$SU(2), which may act
as the non-Abelian part of the standard model gauge group.  SU(3)$\times$SU(2) is the third most common connected two-factor product
group, following SU(2)$\times$SU(2) and $G_2\times$SU(2), which
arise more frequently.  }

\keywords{}

\begin{document}

\maketitle

\flushbottom

\section{Introduction}

F-theory \cite{Vafa-F-theory, Morrison-Vafa-I, Morrison-Vafa-II}
provides a powerful and general nonperturbative approach to the
construction of large classes of string theory vacua.
The construction of an F-theory vacuum 
in $10 - 2 d$ dimensions
depends upon a choice of
a compactification  manifold $B$
that is a complex $d$-fold.   In type IIB language, the axiodilaton of
ten-dimensional supergravity encodes an elliptically fibered
Calabi-Yau $X_{d + 1}$ over the base
$B$.  This Calabi-Yau can have
singularities corresponding to seven-branes carrying gauge groups and
matter in the low-energy $10 - 2 d$ dimensional supergravity theory.

In recent years, by focusing on the geometry of the complex surface
base $B$, a fairly complete global picture of the space of 6d F-theory
constructions and corresponding elliptic Calabi-Yau threefolds has
been developed \cite{KMT-II, clusters, mt-toric, Hodge, Martini-WT,
  Johnson-WT, Wang-WT}.  At a simplified level, the upshot of this
story is that toric bases seem to provide a good representative sample
of the set of all possible base surfaces that support elliptic
Calabi-Yau threefolds, and that an important part of the basic physics
of each base $B$ is captured by the gauge groups and matter in the
``non-Higgsable clusters'' that are present for a generic elliptic
fibration over $B$.

F-theory compactifications to four space-time dimensions on elliptic
Calabi-Yau fourfolds present several additional major complications
and challenges (see {\it e.g.} \cite{Denef-F-theory}), including the
presence of G-flux, which produces a superpotential that lifts some of
the geometric moduli, and world-volume fields on the branes, which can
modify the physics of geometrically non-Higgsable structures.
Nonetheless, at the level of geometry there are many parallels between
the 4d story and the 6d story.  As has been found for CY threefolds,
the range of possible Hodge numbers for known Calabi-Yau fourfolds
seems to be reasonably well captured by elliptic fourfolds over toric
threefold bases.  And also as found for threefolds, geometrically
non-Higgsable gauge groups and matter that are present everywhere in
the Weierstrass moduli space of elliptic fibrations over a given base
$B$ give a strong guide to the physics and and aid in the
classification of 4d F-theory models through the structure of allowed
threefold base geometries.  Non-Higgsable clusters for base threefolds
were studied in specific cases in \cite{Anderson-Taylor, ghst} and
systematically analyzed more generally
in \cite{4d-NHC}.  In \cite{Halverson-WT}, an investigation of the
geometry of one large class of threefold bases was carried out,
focusing on threefolds with the structure of a $\P^1$-bundle over a
complex surface base from among those toric surfaces that themselves
support an elliptic CY threefold.

In this paper we carry out a Monte Carlo study of the set of
toric threefold bases for 4d F-theory models.  The   goal of this
study is to address some basic questions such as:  
\emph{How many toric threefolds
$B$ support an elliptically fibered Calabi-Yau fourfold?
How does the number
of geometrically non-Higgsable gauge group factors grow with $h^{1,
  1} (B)$?}, and: 
\emph{How typical is the product group $SU(3) \times SU(2)$
with jointly charged (quark-like) matter as a subgroup of the
geometrically non-Higgsable gauge group?}  The analysis we perform here
is not intended to give a precise statistical analysis of the physical
distribution of F-theory vacua or even of the full set of fourfolds.
Rather, the goal is to explore a large, potentially characteristic,
set of threefold bases to get a general sense of the scope of the set
of possibilities and  how physical features are distributed.  In particular,
our approach leaves out some classes
of toric threefold bases,  specifically those that cannot be reached by a
sequence of
single blow-ups  and blow-downs from the set of bases connected to $\P^3$.  The
restriction to strictly toric bases also leaves out many bases with
$E_8$ non-Higgsable group factors, which are incorporated more easily
in 6d models through a slight extension of the class of toric bases.
And we do not address here the classification of G-flux, which would
be relevant to analyzing the specific vacua associated with any given
fourfold geometry.

Previous explorations of the range of geometries available for 4d
F-theory vacua have focused on certain simple classes of elliptic
Calabi-Yau fourfolds and threefold bases, in particular threefold
bases that are Fano threefolds
or $\P^1$ bundles over $\P^2$ or other base
surfaces \cite{Klemm-lry,
Mohri,
Berglund-Mayr,
  Grimm-Taylor, Anderson-Taylor, Halverson-WT}.
The set of bases we analyze here includes those bases as special
subsets, though in general the bases explored in the Monte Carlo
analysis have substantially larger values of the Hodge number $h^{1,
  1}(B)$.

The structure of this paper is as follows: In Section \ref{sec:MC}, we
describe the Monte Carlo approach we use to explore threefold bases.
In Section \ref{sec:results}, we give the results of our
investigation, including statistics on typical base geometries and
non-Higgsable clusters.  Some conclusions are contained in Section
\ref{sec:conclusions}.

\section{Monte Carlo on threefold bases}
\label{sec:MC}

\subsection{F-theory basics: bases and gauge groups}

We review here some basic aspects of F-theory, related to the geometry
of the base $B$ and non-Abelian gauge groups associated with Kodaira
singularities in the elliptic fibration.  Much of the basic F-theory
and relevant geometry for describing threefold bases and associated
non-Higgsable clusters is described in \cite{4d-NHC} and
\cite{Halverson-WT}, and  more detail on these subjects can be found in
those papers.  For more general introductions to F-theory, see
\cite{Morrison-TASI, Denef-F-theory, WT-TASI}.

We consider four-dimensional F-theory models that come from an
elliptic fibration with a section over a smooth compact toric threefold
$B$. The total space $X$ is a Calabi-Yau fourfold, which can be
described by a Weierstrass model: \be y^2=x^3+f(s,t,w) x+g(s,t,w) \ee
Here we denote the local coordinates on $B$ by complex variables
$s,t,w$. $f,g$ and the discriminant \be \Delta=4f^3+27g^2 \ee are
sections of line bundles \be
f\in\mc{O}(-4K),\ g\in\mc{O}(-6K),\ \Delta\in\mc{O}(-12K), \ee where
$K$ is the canonical class of the base $B$. The elliptic fiber is
singular at the vanishing locus of $\Delta$. For codimension one loci
on the base, the singularity types were classified by Kodaira
\cite{Kodaira}. In type IIB string theory, there are seven-branes
wrapped on those loci. For different orders of vanishing for $f$ and
$g$, these seven-branes give rise to different non-Abelian gauge
groups in the 4d supergravity theory
\cite{Tate, Bershadsky-all,
Morrison-sn, Grassi-Morrison-2}. We summarize the rules in Table
\ref{t:Kodaira}.

\begin{table}
\begin{center}
\begin{tabular}{|c |c |c |c |c |c |}
\hline
Type &
ord ($f$) &
ord ($g$) &
ord ($\Delta$) &
sing. &  symmetry algebra\\ \hline \hline
$I_0$&$\geq $ 0 & $\geq $ 0 & 0 & none & none \\
$I_n$ &0 & 0 & $n \geq 2$ & $A_{n-1}$ & $\gsu(n)$  or $\gsp(\lfloor
n/2\rfloor)$\\
$II$ & $\geq 1$ & 1 & 2 & none & none \\
$III$ &1 & $\geq 2$ &3 & $A_1$ & $\gsu(2)$ \\
$IV$ & $\geq 2$ & 2 & 4 & $A_2$ & $\gsu(3)$  or $\gsu(2)$\\
$I_0^*$&
$\geq 2$ & $\geq 3$ & $6$ &$D_{4}$ & $\gso(8)$ or $\gso(7)$ or $\gg_2$ \\
$I_n^*$&
2 & 3 & $n \geq 7$ & $D_{n -2}$ & $\gso(2n-4)$  or $\gso(2n -5)$ \\
$IV^*$& $\geq 3$ & 4 & 8 & $\ge_6$ & $\ge_6$  or $\gf_4$\\
$III^*$&3 & $\geq 5$ & 9 & $\ge_7$ & $\ge_7$ \\
$II^*$& $\geq 4$ & 5 & 10 & $\ge_8$ & $\ge_8$ \\
\hline
non-min &$\geq 4$ & $\geq6$ & $\geq12$ & \multicolumn{2}{c|}{does not
appear for SUSY vacua} \\
\hline
\end{tabular}
\end{center}
\caption[x]{\footnotesize  Table of
codimension one
singularity types for elliptic
fibrations and associated non-Abelian symmetry algebras.
In cases where the algebra is not determined uniquely by the orders
of vanishing of $f, g$,
the precise gauge algebra is fixed by monodromy conditions that can be
identified from the form of the Weierstrass model.
}
\label{t:Kodaira}
\end{table}

In general we expand $f$ and $g$
in a local coordinate $w$ near a codimension one locus $w=0$ as follows:
\be
f=f_0+f_1 w+f_2 w^2+f_3 w^3+f_4 w^4+\dots\ ,\ g=g_0+g_1 w+g_2 w^2+g_3 w^3+g_4 w^4+g_5 w^5+\dots\label{expan}
\ee
The coefficients $f_i$, $g_i$ are functions of the other two local coordinates $s$ and $t$.

We consider only generic elliptic fibrations on $B$; that is, we
assume that the functions $f$ and $g$ include all possible
monomials with generic non-vanishing cooefficients. The gauge groups
that arise in this context are called (geometrically) non-Higgsable
gauge groups. Under this condition, not all cases in Table
\ref{t:Kodaira} are relevant in our study. The cases with
ord($\Delta$)$>2\cdot$ord($g$) are excluded. Hence the only possible
non-Higgsable gauge group factors that can arise on a single divisor
(up to possible discrete quotients)
are SU(2), SU(3), $G_2$, SO(7), SO(8), $F_4$, $E_6$, $E_7$ and
$E_8$. There are two different kinds of non-Higgsable SU(2) gauge
groups, corresponding to type $III$ and type $IV$ singular fibers
respectively; we refer to these as  SU(2)${}_{III}$ and $SU(2)_{IV}$.

For the fiber types $IV$, $I_0^*$ and $IV^*$, the gauge group is
specified by additional information encoded in the ``monodromy cover
polynomials'' $\mu(\psi)$ \cite{Bershadsky-all, Morrison-sn,
Grassi-Morrison-2}. Suppose the
divisor is given by a local equation $w=0$, then for the case of type
$IV$,
 \be
 \mu(\psi)=\psi^2-(g/w^2)|_{w=0}=\psi^2-g_2.
 \ee
When $\mu(\psi)$ can be locally factorized into
\be
\mu(\psi)=(\psi+a)(\psi-a),
\ee
the gauge group is SU(3), otherwise it is SU(2). This means that the
gauge group given by a
type $IV$ singular fiber is SU(3) if and only if
$g_2$ is a complete square. The case of type $IV^*$ is similar, where
the monodromy cover polynomial is 
\be
\mu(\psi)=\psi^2-(g/w^4)|_{w=0}=\psi^2-g_4.
\ee
When $g_4$ is a complete square, then the corresponding gauge group is $E_6$, otherwise it is $F_4$. For the case of type $I_0^*$, the monodromy cover polynomial is
\be
\mu(\psi)=\psi^3+(f/w^2)|_{w=0}\psi+(g/w^3)|_{w=0}=\psi^3+f_2\psi+g_3.\label{monocover}
\ee
When $\mu(\psi)$ can be decomposed into three factors:
\be
\mu(\psi)=(\psi+a)(\psi+b)(\psi-a-b),
\ee
the corresponding gauge group is SO(8). Otherwise if it can be decomposed into two factors:
\be
\mu(\psi)=(\psi+a)(\psi^2-a\psi+b),
\ee
the gauge group is SO(7). If it cannot be decomposed at all, then the gauge group is the lowest rank one: $G_2$.

\subsection{Toric bases}

To describe the geometry of the base we will use some basic language
of toric geometry; see \cite{Fulton}. The base three-dimensional 
compact
toric
variety is described by a fan, which is a set of one, two and three
dimensional cones in the integral lattice $N=\mathbb{Z}^3$. The
one-dimensional cones (three-dimensional rays)
$\{v_i=(v_{i,x},v_{i,y},v_{i,z})\in N \}$ correspond to the
toric divisors $\{D_i\}$ on the base ($i=1,\dots,n$). The two-dimensional cone $v_i v_j$ corresponds to the intersection locus of
the two divisors $D_i$ and $D_j$, which is a toric curve. The
three-dimensional cone $v_i v_j v_k$ is the intersection point of
three divisors $D_i$, $D_j$ and $D_k$. There is a universal
requirement on the cones, 
which is that the intersection of any of those cones
is also a cone (or $\varnothing$). Hence, any toric curve $v_i
v_j$ is contained in exactly two three-dimensional cones: $v_i v_j
v_k$ and $v_i v_j v_l$. The essential information about a toric
threefold $B$ is then the set of toric divisors $\{D_i\}$ and the set
of three-dimensional cones $\{\sigma_p=v_i v_j v_k\}$. Furthermore,
the total number of three-dimensional cones is fixed to be
$|\{\sigma_p\}|=2n-4$. For a smooth variety, it is required that the
three-dimensional cones have unit volume: for all $\sigma_p=v_i v_j
v_k$, $|(v_i\times v_j)\cdot v_k|=1$.

Actually, the toric divisors $\{D_i\}$ are not entirely linearly
independent
in homology. There are three linear relations:
\be
\sum_{i=1}^n v_{i,x}D_i=0\ ,\ \sum_{i=1}^n v_{i,y}D_i=0\ ,\ \sum_{i=1}^n v_{i,z}D_i=0.\label{linearrel}
\ee
Hence the rank of the Picard group of $B$ is rk(Pic($B$))$=h^{1,1}(B)=n-3$.

The canonical class $K$ of the base $B$
is given by
\be
K=-\sum_{i=1}^n D_i.
\ee

On $B$ we can define the triple intersection numbers $D_i\cdot
D_j\cdot D_k$. For $i\neq j\neq k$, if $v_i,v_j,v_k$ are in a
three-dimensional cone $\sigma_p$, then $D_i\cdot D_j\cdot
D_k=1$. Otherwise it is zero.  All the triple intersection numbers
involving self products can be determined by the linear relations
(\ref{linearrel}) and the fact that $D_i\cdot D_j\cdot D_k$ vanishes
if $D_i D_j (i\neq j)$ is not a toric curve (Actually those $D_i D_j$
generate the Stanley-Reisner ideal of $B$). However, unlike the case
of 2d toric bases, those triple intersection numbers do not seem to
provide a simple
direct approach to the classification of non-Higgsable
clusters.

Two toric bases $B_1$ and $B_2$ are equivalent
 if their defining rays are related by a lattice isomorphism of
$N=\mathbb{Z}^3$, while keeping the set of cones $\{\sigma_p\}$
unchanged.

\subsection{Blow-ups and blow-downs}

To move between different threefold bases, we can use blow-ups or
blow-downs to increase or decrease the rank of the Picard group.
There are two kinds of blow up operations on toric bases $B$: one can
either blow up a point that corresponds to a three-dimensional cone
$\sigma=v_i v_j v_k$ or blow up a toric curve $v_i v_j$. In the first
case, a new ray $\tilde{v}=v_i+v_j+v_k$ is introduced. The old
three-dimensional cone $\sigma$ is removed, and three new
three-dimensional cones $\tilde{\sigma}_1=v_i v_j \tilde{v}$,
$\tilde{\sigma}_2=v_j v_k\tilde{v}$, $\tilde{\sigma}_3=v_k
v_i\tilde{v}$ are included. For the second case, a new ray
$\tilde{v}=v_i+v_j$ is introduced. Suppose that there are two old 3d
cones $\sigma_1=v_i v_j v_k$ and $\sigma_2=v_i v_j v_l$ that contain
the toric curve $v_i v_j$. They are removed after the blow up. Four
new 3d cones $\tilde{\sigma}_1=v_i v_k\tilde{v}$,
$\tilde{\sigma}_2=v_j v_k\tilde{v}$, $\tilde{\sigma}_3=v_i
v_l\tilde{v}$, $\tilde{\sigma}_4=v_j v_l\tilde{v}$ are included. Note
that $v_i v_j$ is no longer a toric curve after the blow up. Similarly
a blow down is defined to be the inverse process of a blow up (or as
the contraction of a ray). Given a ray $v$, it may or may not be
contracted depending on the neighboring rays. If there are only 3
neighboring rays $v_i,v_j,v_k$ and they satisfy $v=v_i+v_j+v_k$, then
$v$ can be contracted into a point. If there are 4 neighboring rays
$v_i,v_k,v_j,v_l$ (in cyclic order around the curve), if $v=v_i+v_j$
or $v=v_k+v_l$, then $v$ can be contracted into toric curve $v_i v_j$
or $v_k v_l$ respectively. For all the other cases, the ray $v$ cannot
be contracted.

When there are rays $v_i,v_j,v_k,v_l$ that satisfy the relation
$v_i+v_j=v_k+v_l$, and there is a  2d cone $v_iv_j$, then there exists a ``flop'' operation, which is a
combination of a blow up and a blow down; see Figure~\ref{f:flop}.

\begin{figure}
\centering
\includegraphics[height=8cm]{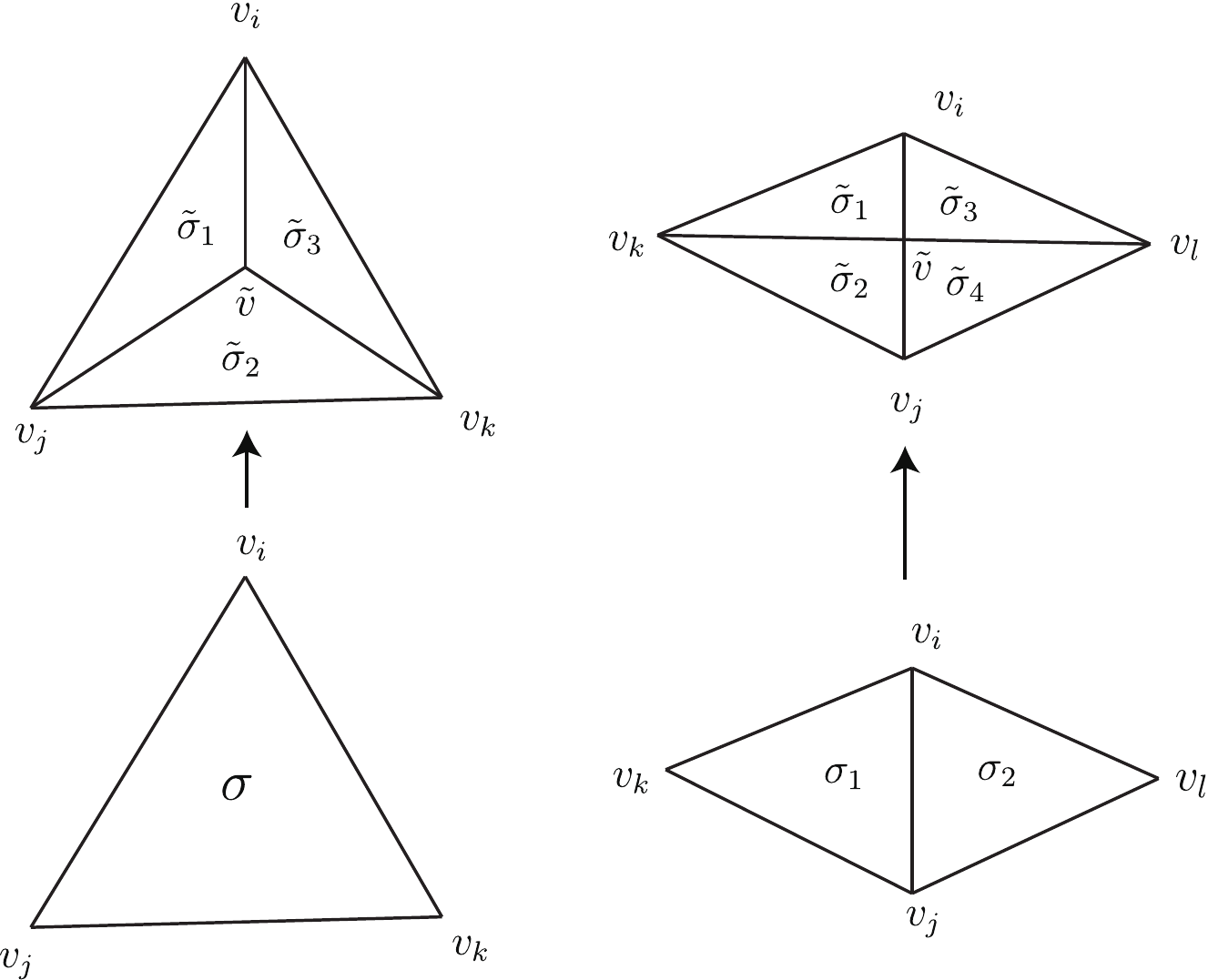}
\caption[x]{\footnotesize Illustration of two different kinds of blow ups, viewed from above. The left case corresponds to blowing up a point $v_i v_j v_k$. The right case corresponds to blowing up a curve $v_i v_j$.}\label{f:blowup}
\end{figure}

\begin{figure}
\centering
\includegraphics[height=8cm]{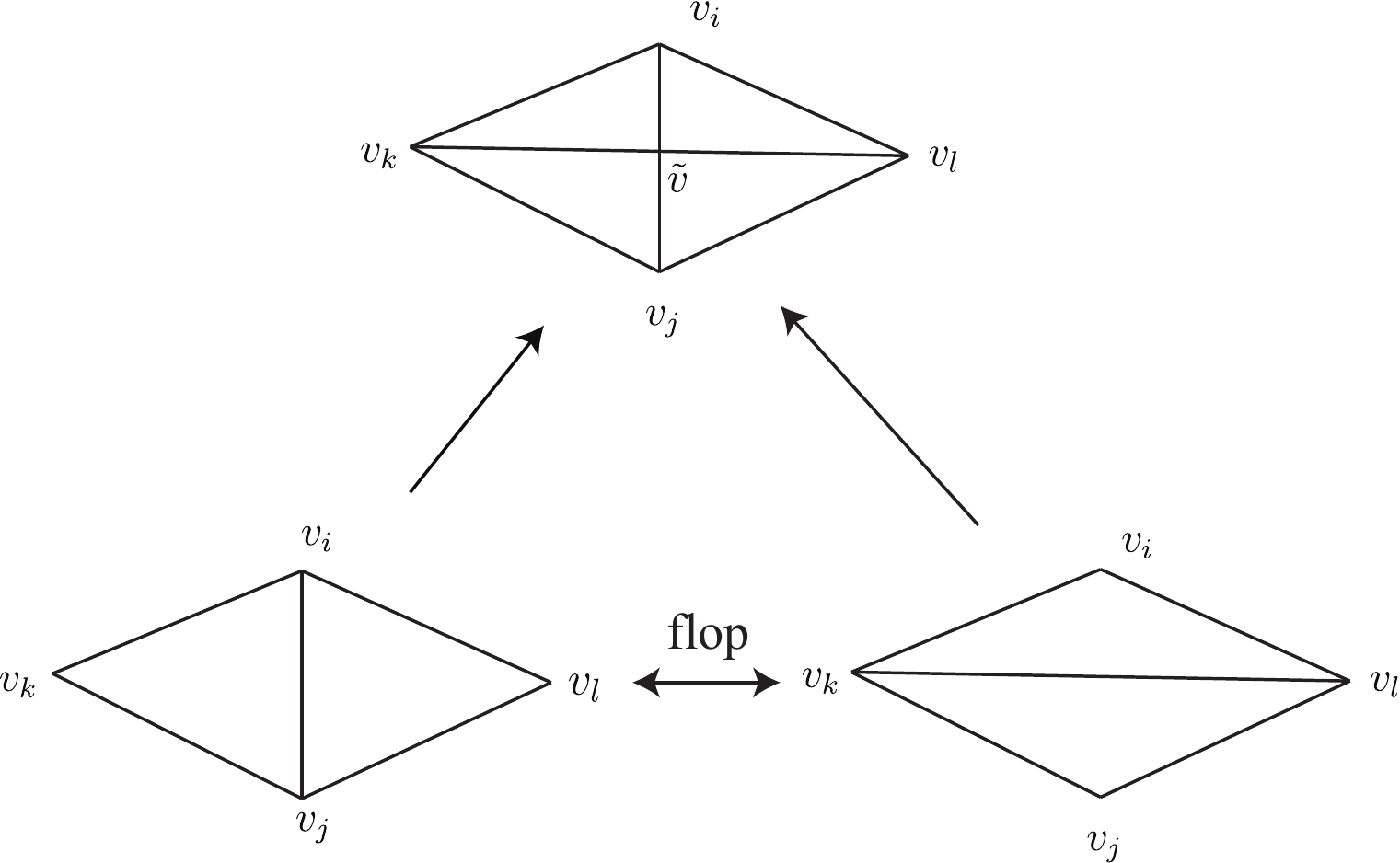}
\caption[x]{\footnotesize Illustration of 
the flop process, which can happen when $v_i+v_j=v_k+v_l$.}\label{f:flop}
\end{figure}

By starting with one toric base, which we take in this work to be
$\P^3$, and performing successive blow-ups and blow-downs, we can
explore a large range of threefold bases that are connected through
these transitions.  We restrict attention to bases that support
elliptic Calabi-Yau fourfolds and associated F-theory models without tensionless
strings,
as determined by the criterion that
there are no codimension 1 or 2 loci where $f, g, \Delta$ vanish to
orders $(4, 6, 12)$.  The precise formulation of this condition for
toric threefold bases is discussed further in \S\ref{sec:toric-monomials}.  While there are allowed toric bases that cannot
be reached from $\P^3$ in this way, the set of bases that are
connected to $\P^3$ by a sequence of single blow-up or blow-down
transitions form a large class of bases; these are the object of study
in this work.  For 2d base surfaces, it is known that all allowed
toric bases are connected through blow-up and blow-down transitions
\cite{mt-toric} to the minimal model bases given by $\P^2$ and the
Hirzebruch surfaces \cite{bhpv, Grassi}; to get to certain bases
(such as, {\it e.g.}, $\F_{12}$) from
a starting point such as $\P^2$, however, requires passing through
intermediate bases that contain points (codimension two loci) where
$f, g$ vanish to orders $(4, 6)$, corresponding to 
tensionless strings and 
superconformal
sectors \cite{Seiberg-Witten, Seiberg-SCFT, Morrison-Vafa-II, SCFT-1,
  SCFT-2}; blowing up these points gives a transition to a base with
an additional tensor multiplet and larger $h^{1, 1}$.  We do not
include bases containing such codimension two loci in our study; as in
the 6d (complex surface base) case this means that certain toric
threefold bases will be disconnected from the set under study.
Furthermore, the analogue (described using Mori theory \cite{Mori}) of
the minimal base surfaces for threefold bases is not known explicitly,
and there is no proof that all threefold bases that support elliptic
Calabi-Yau threefolds are connected through simple geometric
transitions.  Nonetheless, the class of bases that we study here,
which are connected to $\P^3$ by a sequence of allowed blow-up or
blow-down transitions through allowed bases that have no codimension
two $(4, 6)$ loci, seem to form a large and sufficiently diverse class
of bases to give interesting information about a fairly generic class
of threefold F-theory bases.

\subsection{Toric monomials and non-Higgsable clusters}
\label{sec:toric-monomials}

The origin $(0,0,0)\in N$ represents a complex torus
$(\mathbb{C}^*)^3$, which has three coordinates $S,T,W$. The sets of
holomorphic monomials that are sections of specific line bundles
 on $B$ are subsets of the dual integral lattice
$M=$ Hom$(N,\mathbb{Z})$. The monomial $S^a T^b W^c$ is represented by
$(a,b,c)\in M$. Each of the three-dimensional cones $\sigma=v_i v_j
v_k$ represents a local coordinate patch. Over this patch the ring of
holomorphic monomials is described by the dual cone
\be
\sigma^*=\{u\in M|\langle u,v_i\rangle\geq 0\ ,\ \langle u,v_j\rangle\geq 0\ ,\ \langle u,v_k\rangle\geq 0\}.
\ee
Using this information we can write down the local coordinates in that
patch and the transition rules between different patches. If the three
divisors $D_i$, $D_j$ and $D_k$ are given by local equations $w=0$,
$s=0$ and $t=0$, then the monomial $u\in M$ corresponds to $w^{\langle
  u, v_i\rangle} s^{\langle u, v_j\rangle} t^{\langle u,v_k\rangle}$.

Now we can construct the set of monomials appearing in $f$, $g$ and $\Delta$. We denote them by $\{f\}$, $\{g\}$ and $\{\Delta\}$. They are given by:
\be
\bsp
&\{f\}=\{u\in M|\forall v_i\ ,\ \langle u,v_i\rangle\geq -4\},\\
&\{g\}=\{u\in M|\forall v_i\ ,\ \langle u,v_i\rangle\geq -6\},\\ 
&\{\Delta\}=\{u\in M|\forall v_i\ ,\ \langle u,v_i\rangle\geq -12\}.
\end{split}
\ee
The order of vanishing of $f$, $g$ and $\Delta$ on a toric divisor $D_i$ is then
\be
\bsp
&\text{ord}_{D_i}(f)=\min(\langle u,v_i\rangle+4)|_{u\in\{f\}},\\
&\text{ord}_{D_i}(g)=\min(\langle u,v_i\rangle+6)|_{u\in\{g\}},\\
&\text{ord}_{D_i}(\Delta)=\min(\langle u,v_i\rangle+12)|_{u\in\{\Delta\}}.
\end{split}
\ee
We can also write down the order of vanishing of $f$, $g$ and $\Delta$
on a toric curve $D_i D_j$:
\be
\bsp
&\text{ord}_{D_i D_j}(f)=\min(\langle u,v_i\rangle+\langle u,v_j\rangle+8)|_{u\in\{f\}},\\
&\text{ord}_{D_i D_j}(g)=\min(\langle u,v_i\rangle+\langle u,v_j\rangle+12)|_{u\in\{g\}},\\
&\text{ord}_{D_i D_j}(\Delta)=\min(\langle u,v_i\rangle+\langle u,v_j\rangle+24)|_{u\in\{\Delta\}}.
\end{split}
\ee
$\text{ord}_{D_i D_j}(f,g,\Delta)$ can be greater or equal to
$\text{ord}_{D_i}(f,g,\Delta)+\text{ord}_{D_j}(f,g,\Delta)$. This is
different from the case of 2d bases, where the order of vanishing on
the intersection point of two divisors is always the sum of the orders
of vanishing on those two divisors.

Similarly we can write down the order of vanishing of $f$, $g$ and $\Delta$ on the point $D_i D_j D_k$:
\be
\bsp
&\text{ord}_{D_i D_j D_k}(f)=\min(\langle u,v_i\rangle+\langle u,v_j\rangle+\langle u,v_k\rangle+12)|_{u\in\{f\}},\\
&\text{ord}_{D_i D_j D_k}(g)=\min(\langle u,v_i\rangle+\langle u,v_j\rangle+\langle u,v_k\rangle+18)|_{u\in\{g\}},\\
&\text{ord}_{D_i D_j D_k}(\Delta)=\min(\langle u,v_i\rangle+\langle u,v_j\rangle+\langle u,v_k\rangle+36)|_{u\in\{\Delta\}}.
\end{split}
\ee

For a good F-theory base, we exclude the cases where the order of vanishing of $f$ and $g$ reaches $4$ and $6$ on a divisor:
\be
\text{ord}_{D_i}(f)\geq 4\ ,\ \text{ord}_{D_i}(g)\geq 6
\; \Rightarrow \; {\rm excluded}
\label{eq:excluded-1}
\ee
or on a curve:
\be
\text{ord}_{D_i D_j}(f)\geq 4\ ,\ \text{ord}_{D_i D_j}(g)\geq 6
\; \Rightarrow \; {\rm excluded}
\ee
As mentioned above, in the case of a curve a $(4, 6)$ locus indicates
the appearance of
a  tensionless string, so that the low-energy theory does not have a
conventional field theory description; such curves can be blown up to get
another base that generally supports a less singular elliptic CY threefold.
We also exclude the bases where the order of vanishing of $f$ and $g$
reaches $8$ and $12$ on a toric intersection point\footnote{It is not
  clear whether it is problematic or not when the orders of vanishing of
  $f$ and $g$ reach (4,6) at a codimension-three locus
\cite{codimension-3}. Here, as in
\cite{4d-NHC}, we use
  the weaker condition (\ref{eq:excluded-8-12}).}:
\be
\text{ord}_{D_i D_j D_k}(f)\geq 8\ ,\ \text{ord}_{D_i D_j D_k}(g)\geq
12
\; \Rightarrow \; {\rm excluded}
\label{eq:excluded-8-12}
\ee
Practically, since the elliptic fibration is generic we only need to
check the order of vanishing for $g$. 

Apart from  the constraints
(\ref{eq:excluded-1})--(\ref{eq:excluded-8-12}), there are also other cases
where $\text{ord}_{D_i}(f)\geq 4$, $\text{ord}_{D_i}(g)=5$, but the
expansion coefficient $g_5$ in (\ref{expan}) contains more than one
monomial. When this happens, $g_5$ is not constant, which means
that there
is a locus $\Sigma\in D_i$ such that $\text{ord}_{\Sigma}(g)=6$. This
type of  (4, 6) singularity on a curve
is analogous to that which arises at points on 
the (-9), (-10), (-11)-curves on
2d bases \cite{clusters}.  
We exclude toric bases with such $(4, 6)$ curves, though they may
admit blow-ups to non-toric bases that support good F-theory
compactifications.

Using the information of $\{f\}$ and $\{g\}$, we can read off the
non-Higgsable
gauge group on each divisor. We explicitly 
describe the rules for
non-Higgsable
 type
$IV$, $IV^*$ and $I_0^*$ singularities; 
see also
\cite{Anderson-Taylor, 4d-NHC}.

When $\text{ord}_{D_i}(f)\geq 2$, $\text{ord}_{D_i}(g)=2$, the
singularity type is $IV$. In this case when $g_2$ only contains one
monomial $u\in M$, and
furthermore $u \in 2M$ is even and therefore a perfect square, then the gauge group is SU(3). Otherwise the gauge group is type  SU(2)$_{IV}$. 

When $\text{ord}_{D_i}(f)\geq 3$, $\text{ord}_{D_i}(g)=4$, the
singularity type is $IV^*$. In this case when $g_4$ only contains one
monomial $u\in M$, and
furthermore $u \in 2M$, then the gauge
group is $E_6$. Otherwise the gauge group is $F_4$.

 When $\text{ord}_{D_i}(f)\geq 2$, $\text{ord}_{D_i}(g)=3$ or $\text{ord}_{D_i}(f)=2$, $\text{ord}_{D_i}(g)>3$, the singularity type is $I_0^*$. When the monodromy cover polynomial (\ref{monocover}) can be written as
 \be
 \mu(\psi)=\psi^3+f_2\psi+g_3=\psi^3-(a^2-b)\psi+ab,\label{monoso7}
 \ee
 the gauge group is SO(7) or SO(8). The gauge group is SO(8) only when $\mu(\psi)$ can be written as
 \be
 \mu(\psi)=\psi^3+f_2\psi+g_3=\psi^3-(a^2+ab+b^2) \psi-(a+b)ab.\label{monoso8}
 \ee
Now if $\text{ord}_{D_i}(f)=2$ and $\text{ord}_{D_i}(g)>3$,
  $\mu(\psi)=\psi^3+f_2\psi$, the gauge group is either SO(7) or
  SO(8). The gauge group is SO(8) only when $a=0$ or $b=0$ or $a+b=0$
  in (\ref{monoso8}). Then it is required that $f_2$ only contains one
  monomial $u\in M$, and
furthermore $u \in 2M$. Otherwise the gauge group is SO(7).

For the other case, $\text{ord}_{D_i}(f)\geq 2$ and
$\text{ord}_{D_i}(g)=3$,
 for generic $f, g$, $\mu$ can only have the form
(\ref{monoso7}) if $f_2$ and $g_3$ each contain only single monomials
 and $b \sim a^2$.  This can be seen by simply considering the number
 of independent monomials in $a, b$ compared to $f_2, g_3$.
Thus, when $f_2 \sim a^2$ and  $g_3 \sim a^3$ for a single monomial
$a$,
$\mu(\psi)$ can be written in form of
(\ref{monoso8}) and the gauge group is SO(8); otherwise the gauge
group is $G_2$.

\subsection{Fourfold geometry}

Much of the relevant geometry of the generic
elliptic Calabi-Yau fourfold fibered over the threefold
base $B$ can be read off directly from the geometry of $B$.

From the full non-Abelian non-Higgsable gauge group $G$, we can compute
the Hodge number $h^{1,1}(X)$ of the Calabi-Yau fourfold, using the
Shioda-Tate-Wazir formula \cite{Morrison-Vafa-II, stw}
\be
h^{1,1}(X)
\cong \tilde{h}^{1, 1}(X)=h^{1,1}(B)+\text{rk}(G)+1.
\ee
Here we assume that there is no non-Higgsable U(1) gauge group. For
all the 2d toric bases, such contributions to $h^{1, 1}(X_3)$ never
appear \cite{mt-toric}, though we do not know for sure that this
cannot happen for toric 3d bases.

We can also compute the Hodge number $h^{3,1}(X)$, using an approximate Batyrev type formula \cite{Hodge-btw}:
\begin{eqnarray}
h^{3,1}(X) &\cong &
\tilde{h}^{3, 1}(X)\\
&= &|\{f\}|+|\{g\}|-\sum_{\Theta\in\Delta,\dim\Theta=2}l'(\Theta)-4+\sum_{\Theta_i\in\Delta,\Theta_i^*\in\Delta^*, \dim(\Theta_i)=\dim(\Theta_i^*)=1}l'(\Theta_i)\cdot l'(\Theta_i^*)\,.\nonumber
\end{eqnarray}
Here $\Delta^*$ is the convex hull of $\{v_i\}$ and $\Delta$ is the dual polytope of $\Delta^*$, defined to be
\be
\Delta=\{u\in\mathbb{R}^3|\forall v\in\Delta^*\ ,\ \langle u,v\rangle\geq -1\}.
\ee
The symbol $\Theta$ denotes 2d faces of $\Delta$. $\Theta_i$ and
$\Theta_i^*$ denote the 1d edges of the polytopes $\Delta$ and
$\Delta^*$. $l'(\cdot)$ counts the number of integral interior points
on a face.
While this formula is only proven for a subclass of toric threefold
bases, and may be off by small amounts for some choices of $B$, we
expect that it is a good approximate measure of $h^{3, 1} (X)$.

Because both formulae used here are not rigorously proven for all
toric threefold bases, and we only use them as approximate measures of
the Hodge numbers, we have used tildes to denote the approximate Hodge
numbers $\tilde{h}^{1, 1}, \tilde{h}^{3, 1}$ given by these formulae.

\subsection{Random walks on the connected set of toric threefold bases}
\label{sec:random}

In order to characterize generic properties of a 3d toric base in 4d
F-theory, we can perform a random walk from some starting point, say
$\mathbb{P}^3$. In each step of the random walk, the base may be blown
up or blown down to get another acceptable base.  
Depending upon the
weighting of the probabilities of each move, we get a specific resulting
distribution on the set
${\cal C}$ of connected valid 3d toric bases without $(4,
6)$ curves.

We perform the random walk using an equal weighting for each valid
blow-up or blow-down from a given base $B \in{\cal C}$.  It is easy to see that a
random walk on a graph where each node $V_i$ has $n_i$ neighbors,
where neighbors are chosen uniformly on each step of the walk, will
give a distribution on nodes proportional to $n_i$, since the
probability of traversing each link in either direction is then equal.
A potential obstruction to using this algorithm in the case at hand is
the computational burden of determining which neighbors are valid; for
a toric threefold base having a fan with $n$ rays and $h^{1, 1}
(B)=n-3$, the number of faces (three-dimensional cones) is $2n-4$, and
the number of edges (two-dimensional cones) is is $3n-6$, so the
number of possible moves goes as $6n$.  For $n \sim 100$, this makes
it very costly to evaluate all possible blow-ups and blow-downs for
validity.  Thus, we use a simpler algorithm of simply choosing a
possible blow-up or blow-down at random from the set of all the
$6n-10$ possible 3d cones, 2d cones, and rays, and then testing the
chosen move to see if it results in an allowed base.  If the tested
step does not lead to an allowed base, we try again.  This effectively
gives a random walk where all allowed moves are weighted equally, so
that over a large number of steps we expect a ``thermal'' distribution
in which the probability of each base $B$ in the set
${\cal C}$ connected to the
starting base $\P^3$ is proportional to $n_i$, the number of valid
neighbors to which $B$ is connected by single blow-up or blow-down
moves.  To get a uniform distribution on the set of allowed bases we
need
 to weight our statistics by the factor $1/n_i$
for each base.  Because we do not explicitly compute $n_i$ for the
reasons given above, we estimate the weighting factor in a crude way
by computing the number $t$ of tries needed to identify an allowed
neighbor. Naively the number of allowed neighbors of a given base $B$ should
then be $(6n-10)/\langle t \rangle$, where $\langle t \rangle$ is the
average number of tries needed to identify an allowed neighbor over
many trials on the base $B$.  The
weighting factor $1/n_i$ can therefore be estimated as
$\langle t \rangle/(6n-10)$, so we can estimate quantities
statistically by weighting each base with the factor $t/(6n-10)$.

Sometimes, however, the different neighbors of one base can be equivalent. For
example, consider a graph with only three nodes: $\P^3$,
blp$_{cone}\P^3$ and blp$_{curve}\P^3$. blp$_{cone}\P^3$ and
blp$_{curve}\P^3$ denote the bases that result from blowing up a
(3d) cone
or a curve on $\P^3$ respectively. We explicitly list the rays and
3d cones for these three toric threefold bases below:

\be
\bsp
\P^3\ :\ &v_1=(1,0,0),v_2=(0,1,0),v_3=(0,0,1),v_4=(-1,-1,-1),\\&\{\sigma_p\}=\{v_1 v_2 v_3,v_1 v_2 v_4,v_1 v_3 v_4, v_2 v_3 v_4\}
\end{split}
\ee

\be
\bsp
\text{blp}_{cone}\P^3\ :
\ &v_1=(1,0,0),v_2=(0,1,0),v_3=(0,0,1),v_4=(-1,-1,-1),v_5=(1,1,1),\\&\{\sigma_p\}=\{v_1
v_2 v_4,v_1 v_3 v_4, v_2 v_3 v_4,v_1 v_2 v_5,v_1 v_3 v_5,v_2 v_3 v_5\} 
\end{split}
\ee

\be
\bsp
\text{blp}_{curve}\P^3\ :\ &
v_1=(1,0,0),v_2=(0,1,0),v_3=(0,0,1),v_4=(-1,-1,-1),v_5=(1,1,0),\\&
\{\sigma_p\}=\{v_1 v_3 v_4, v_2 v_3 v_4,v_1 v_3 v_5,v_1 v_4 v_5,v_2
v_3 v_5,v_2 v_4 v_5\} 
\end{split}
\ee

There are four ways to get blp$_{cone}\P^3$ and six ways to get
blp$_{curve}\P^3$
from blowing up a cone or curve on $\P^3$, since there are 4 3d-cones and 6 2d-cones in the
toric fan of $\P^3$. This means that naively $\P^3$ has 10 neighbors,
and the base is weighted by $1/10$. Now, if we perform a random walk on
this graph, the expected probability ratio is
$p(\P^3):p($blp$_{cone}\P^3):p($blp$_{curve}\P^3)=10:4:6$. Then after
we weight $p(\P^3)$ by a factor $1/10$, the expected probability ratio
becomes $1:4:6$, which is still far from uniform. To fix this problem
we compute the symmetry factor $F$ of each base, which is defined to be
the order of the
subgroup of the permutation group acting on the toric divisors of
the base that preserves the cone structure. For example, for the base $\P^3$, since all the four rays
$v_1,v_2,v_3,v_4$ 
 can be permuted arbitrarily without changing the cone structure,
$F(\P^3)=24$. For the base 
$\text{blp}_{cone}\P^3$, the divisors corresponding to $v_1,v_2$ and
$v_3$ 
 can be permuted, hence $F(\text{blp}_{cone}\P^3)=6$. For the base
$\text{blp}_{curve}\P^3$, there are two
symmetric divisor pairs:
$(v_1,v_2)$ and $(v_3,v_4)$, hence
$F(\text{blp}_{curve}\P^3)=4$. After we multiply those symmetry
factors by the ratio $1:4:6$, then we achieve a uniform
distribution. In general, if there are $m$ ways to get equivalent bases
$B$ from $k$ equivalent bases $A$ with symmetry factor $F(A)$, then the symmetry
factor of $B$ satisfies $mF(B)=kF(A)$. Hence this inclusion
of symmetry factors solves the problem.

For a general base with a large number of rays, the probability of
having
a nontrivial symmetry is negligible. So practically, the
inclusion of symmetry factors only affects the statistics of bases
with a
number of rays $n\lesssim 10$.

In the following section we present results of this Monte Carlo approach.

\section{Results} 
\label{sec:results}

\subsection{Choices of Monte Carlo parameters}

Our primary analysis was carried out by doing 100 independent runs of
100,000 bases, each starting at $\P^3$ and exploring a subset of the
bases in the connected set ${\cal C}$ using a random walk as described
above. In the remaining parts of this paper we
refer to these 100 runs as ``unbounded''
runs, to distinguish them from other runs (with bounded $h^{1,1}(B)$)
described in \S\ref{sec:number}.  The first 500 or so bases in each
unbounded run had atypically small values
of $h^{1, 1}(B)$.  We compute our statistics based on the subset after
each run has approximately thermalized, by dropping the first 1000 bases.
As we discuss further in the following sections, each run rapidly
seems to have entered a local region, or {\it domain}, of the allowed
space of bases that may only be connected to the other domains through
relatively rare paths in the graph that may require an excursion to
relatively low $h^{1, 1}$.  Thus, it seems that these individual runs
are not truly thermalized in a global sense.  Nonetheless, the
distributions in each domain are sufficiently similar and regularly
distributed between domains that we take the set of data from the 100
independent runs as presumably relatively representative of the full
set ${\cal C}$ of connected bases.  Further more extensive analysis
would be necessary to rigorously demonstrate or counter this
hypothesis.  We also note in parts of the analysis some distinctions
between the different domains explored by the independent runs.

\subsection{Distribution and number of threefold bases}

\subsubsection{Distribution of bases}

We begin by considering the distribution of bases as estimated by the
Monte Carlo analysis using some fairly simple measures.

As described in \S\ref{sec:random}, we estimate the proper
weighting factor for each base $B$ encountered by $t\cdot
F(B)/(6n-10)$, where $n$ is the number of rays for the fan of $B$, $t$
is the number of tries needed to find an allowed neighbor, and $F (B)$
is the symmetry factor of $B$.

\begin{figure}
\begin{center}
\includegraphics[height=10cm]{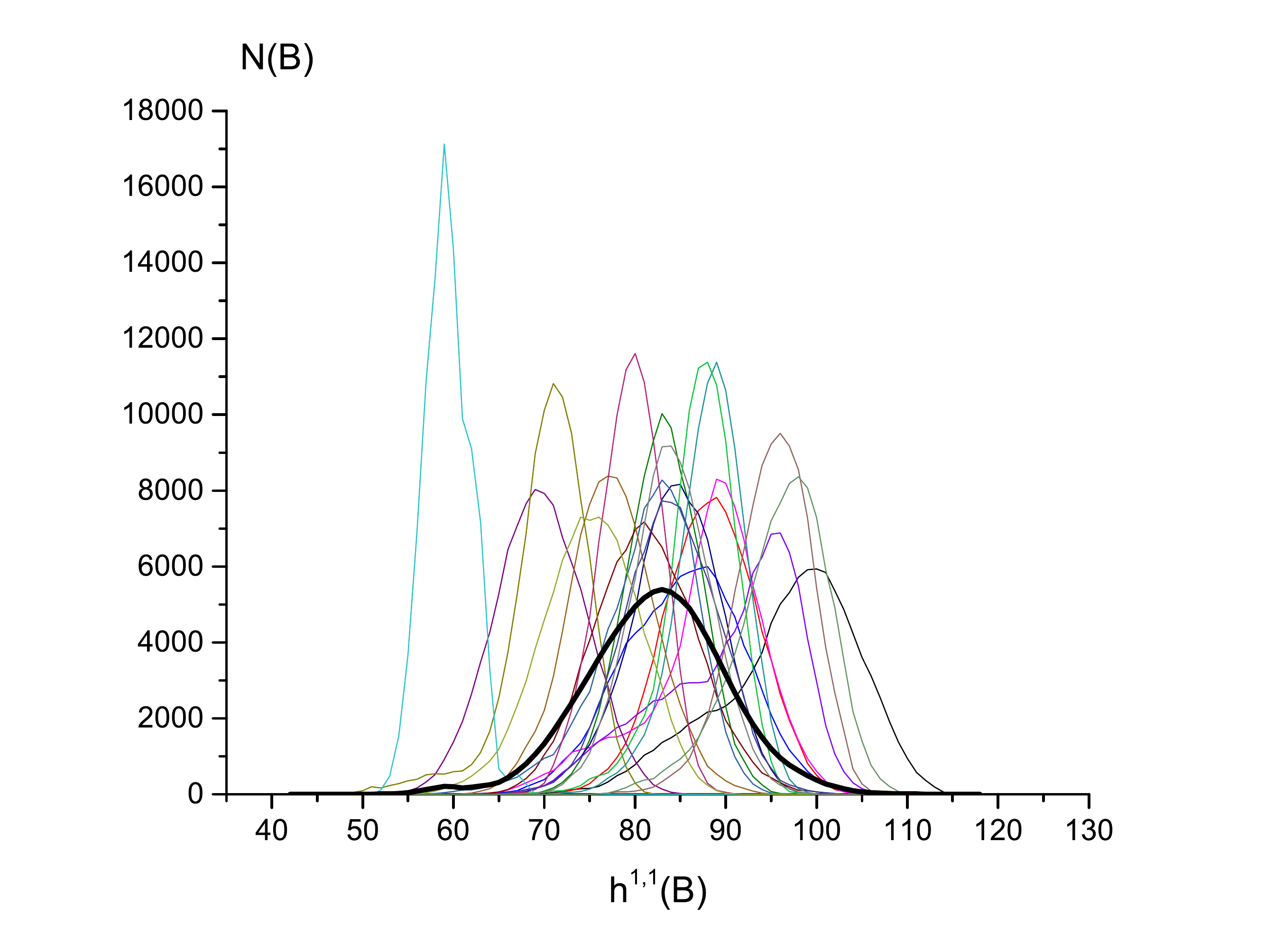}
\end{center}
\caption[x]{\footnotesize  The distribution of $h^{1, 1}(B)$,
  including weighting factors. Bold black curve is total 
  distribution (divided by the number of runs, 100), colored curves
  are some example distributions from 
  individual runs. The total number of samples in each run is normalized to 100,000.}
\label{f:h11-distribution}
\end{figure}

The total distribution is graphed in
Figure~\ref{f:h11-distribution} and compared to the distributions for
several individual runs.

From the distinct shapes of the distributions from individual runs, we
see evidence for the observation mentioned above that each run is
probing a different domain in the connected space of bases.  One
particular run, for example, probed a set of bases with anomalously
low $h^{1,1}(B)$ values, giving rise to the small bump on the left
tail of the total distribution shown in
Figure~\ref{f:h11-distribution}.  A much longer (perhaps impossibly
long in practice) Monte Carlo run would be needed to have true
thermalization between the domains in a single run.

The overall mean and standard deviation of $h^{1, 1}(B)$ are
\begin{eqnarray}
\langle h^{1, 1} (B) \rangle & = &  { 82}\\
\sigma (h^{1, 1}(B)) & = &  { 6} \,.
\end{eqnarray}
Comparing the distinct runs,
the mean value in each run ranged from 59.4 to
96.8, with a standard deviation of 6.

As mentioned in the introduction, simple bases such as toric Fano
threefolds, $\P^1$ bundles over $\P^2$ and $\P^2$ bundles over $\P^1$,
as studied in \cite{Klemm-lry, Mohri, Berglund-Mayr},
have very small values of $h^{1, 1}(B)$, and are only encountered in
the first stages of the Monte Carlo runs, before thermalization.
Larger classes of $\P^1$ bundles over more general base surfaces were
explored in \cite{Anderson-Taylor} and \cite{Halverson-WT}.  In
particular, in \cite{Halverson-WT}, the full set of threefolds that
have the form of $\P^1$ bundles over toric surfaces $S$ that
themselves support elliptic Calabi-Yau threefolds was explored.  That
set included threefolds with a larger range of values of $h^{1,
  1}(B)$, and is more closely analogous to the distribution of bases
studied here.  As we mention again below, the qualitative distribution
of physical features on that set is roughly
compatible with what we have found
in the Monte Carlo analysis, although the $\P^1$-bundle threefolds
have certain characteristic features that affect the distribution of
{\it e.g.} non-Higgsable clusters that arise on those bases.

\begin{figure}
\begin{center}
\includegraphics[height=8cm]{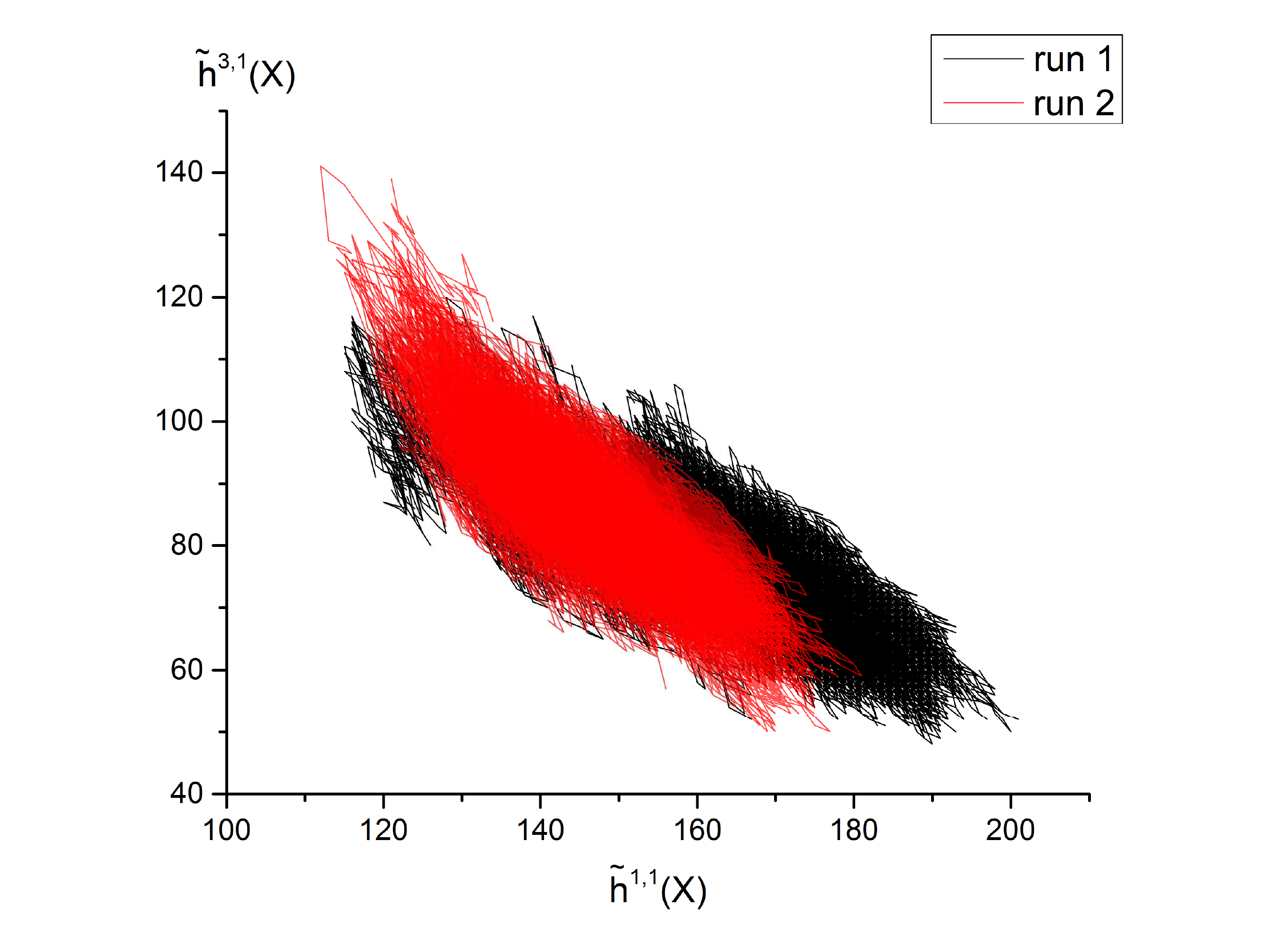}
\end{center}
\caption[x]{\footnotesize  (Approximate)
Hodge numbers for generic elliptically fibered fourfolds over the
bases $B$ encountered in two of the
  random walks through the space of connected bases.}
\label{f:sample-Hodge}
\end{figure}

\begin{figure}
\begin{center}
\includegraphics[height=8cm]{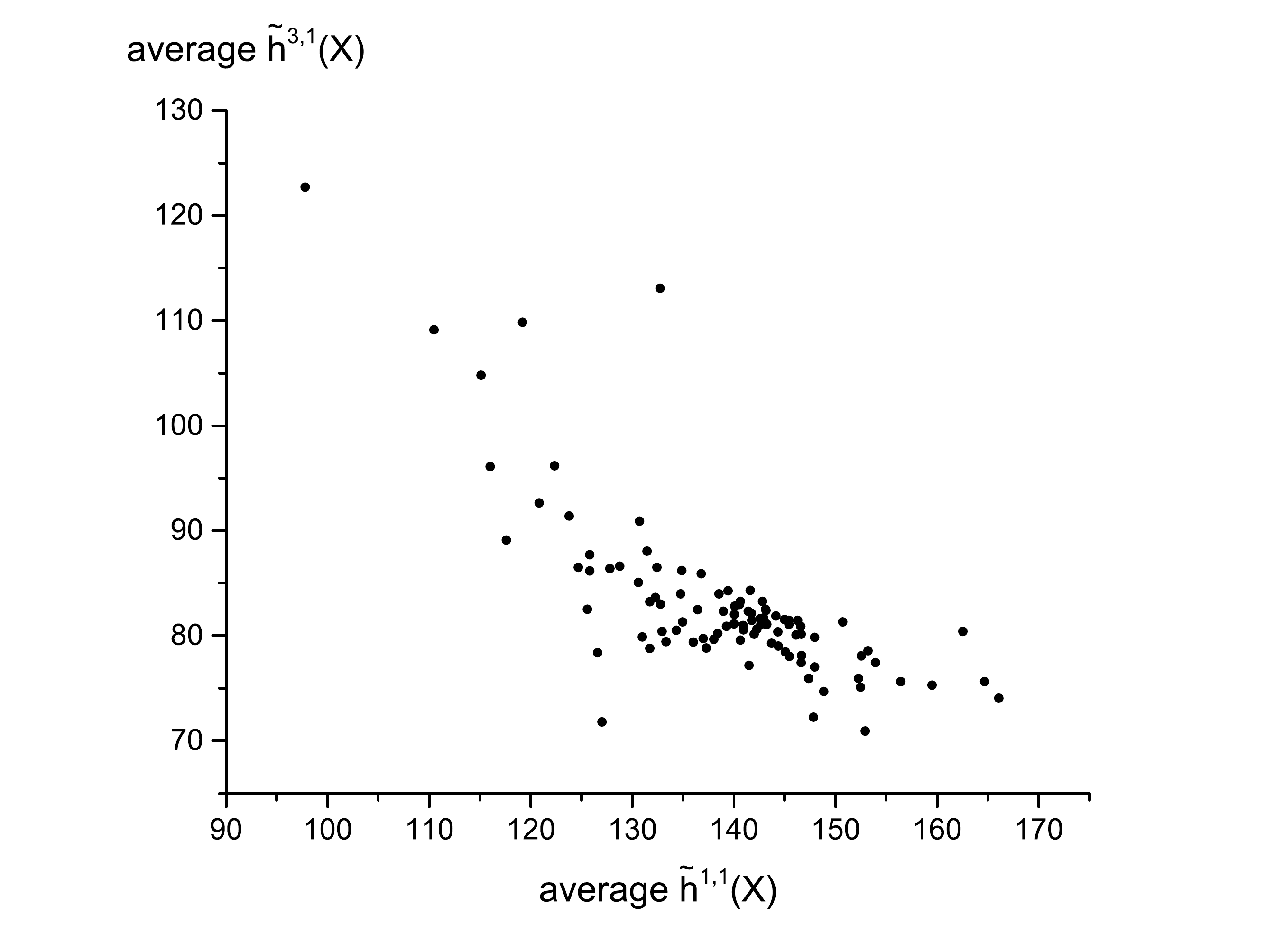}
\end{center}
\caption[x]{\footnotesize  
Mean values of the
(approximate)
Hodge numbers
for generic elliptic  Calabi-Yau fourfolds
 over the threefold bases
 encountered in each of the independent Monte Carlo
runs.}
\label{f:mean-Hodge}
\end{figure}

\begin{figure}
\begin{center}
\includegraphics[height=10cm]{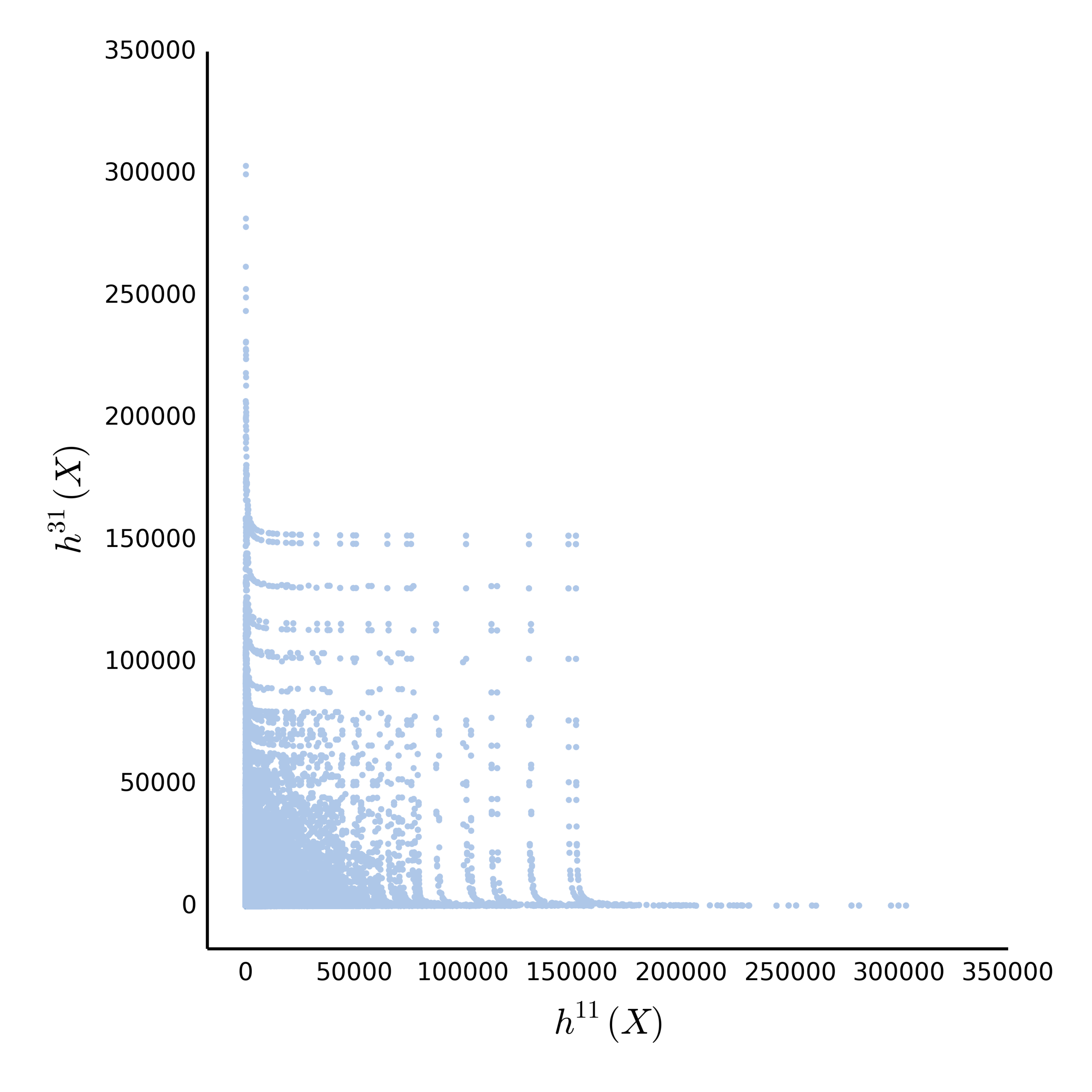}
\end{center}
\caption[x]{\footnotesize   Distribution of Hodge numbers for
  Calabi-Yau fourfolds constructed as hypersurfaces in weighted
  projective space using reflexive polytopes
  \cite{Kreuzer-Skarke-4d}.}
\label{f:Kreuzer-4d-Hodge}
\end{figure}

It is helpful to get a sense of how each run explores the space of
possible bases by graphing the set of (approximate) Hodge numbers
$\tilde{h}^{1, 1} (X), \tilde{h}^{3, 1} (X)$ of the generic elliptically
fibered Calabi-Yau fourfolds over the sets of bases $B$
explored by the separate runs.
Two sample runs are shown in Figure~\ref{f:sample-Hodge}.  The
mean Hodge numbers across each run are shown in
Figure~\ref{f:mean-Hodge}.
These Hodge numbers can be compared to the distribution of
Hodge numbers known for general Calabi-Yau fourfolds constructed using
toric and related methods \cite{Klemm-lry, lsw-4d,
Kreuzer-Skarke-4d}, as depicted in
Figure~\ref{f:Kreuzer-4d-Hodge}.

\begin{figure}
\begin{center}
\includegraphics[height=8cm]{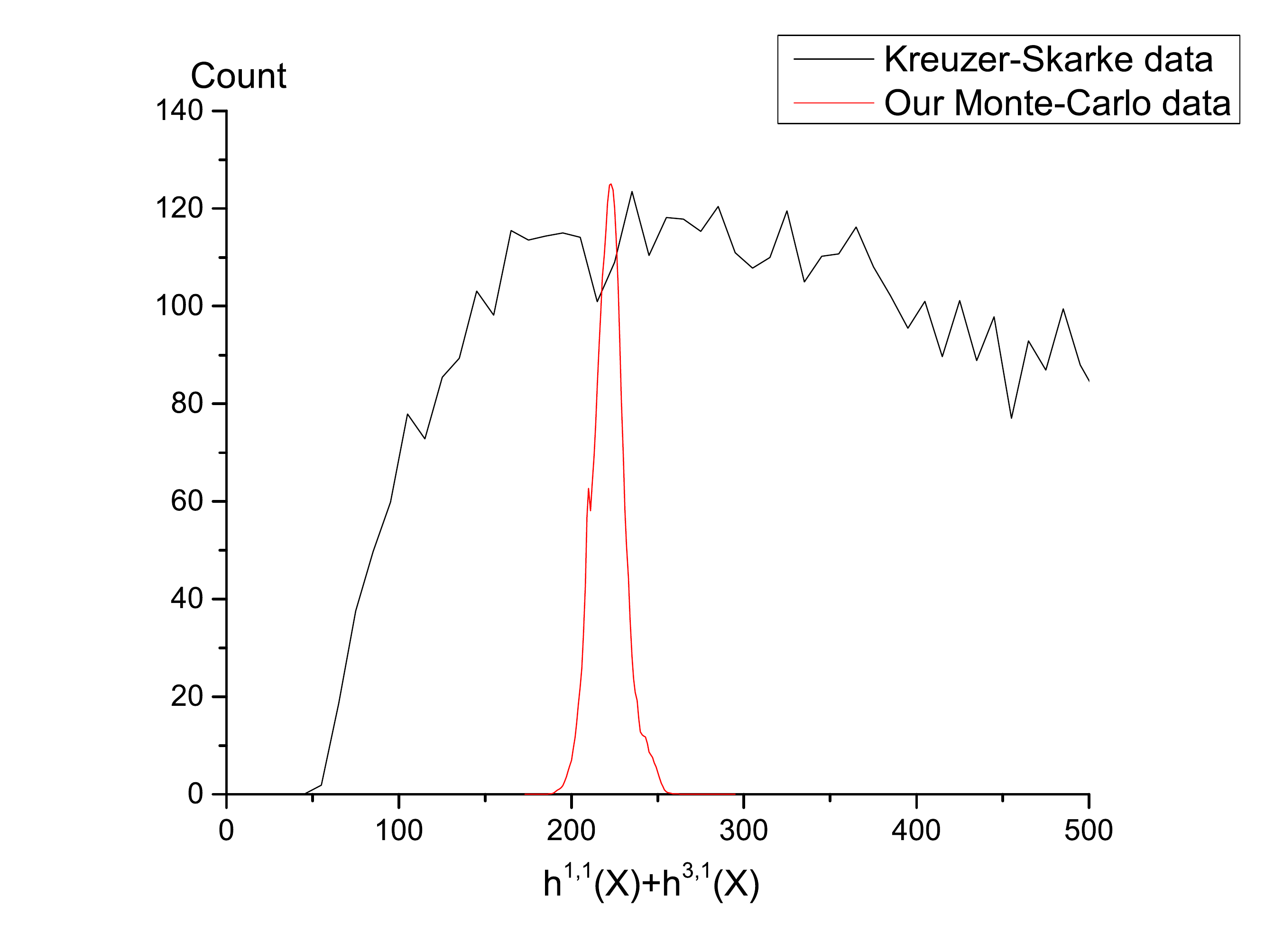}
\end{center}
\caption[x]{\footnotesize  The weighted distribution of bases found in
the Monte Carlo as a function of
$\tilde{h}^{1, 1} (X) +\tilde{h}^{3, 1} (X)$, compared to the set of
fourfolds identified in \cite{lsw-4d}, as hypersurfaces in 5d reflexive polytopes corresponding to transverse weight systems.
The scale on the Monte Carlo data is not significant and is chosen
based on the size of the Monte Carlo runs, which happens to give it a
compatible scale for easy comparison with the
fourfold data (available from \cite{Kreuzer-Skarke-data}).
}
\label{f:compare-ks}
\end{figure}

\begin{figure}
\begin{center}
\includegraphics[height=8cm]{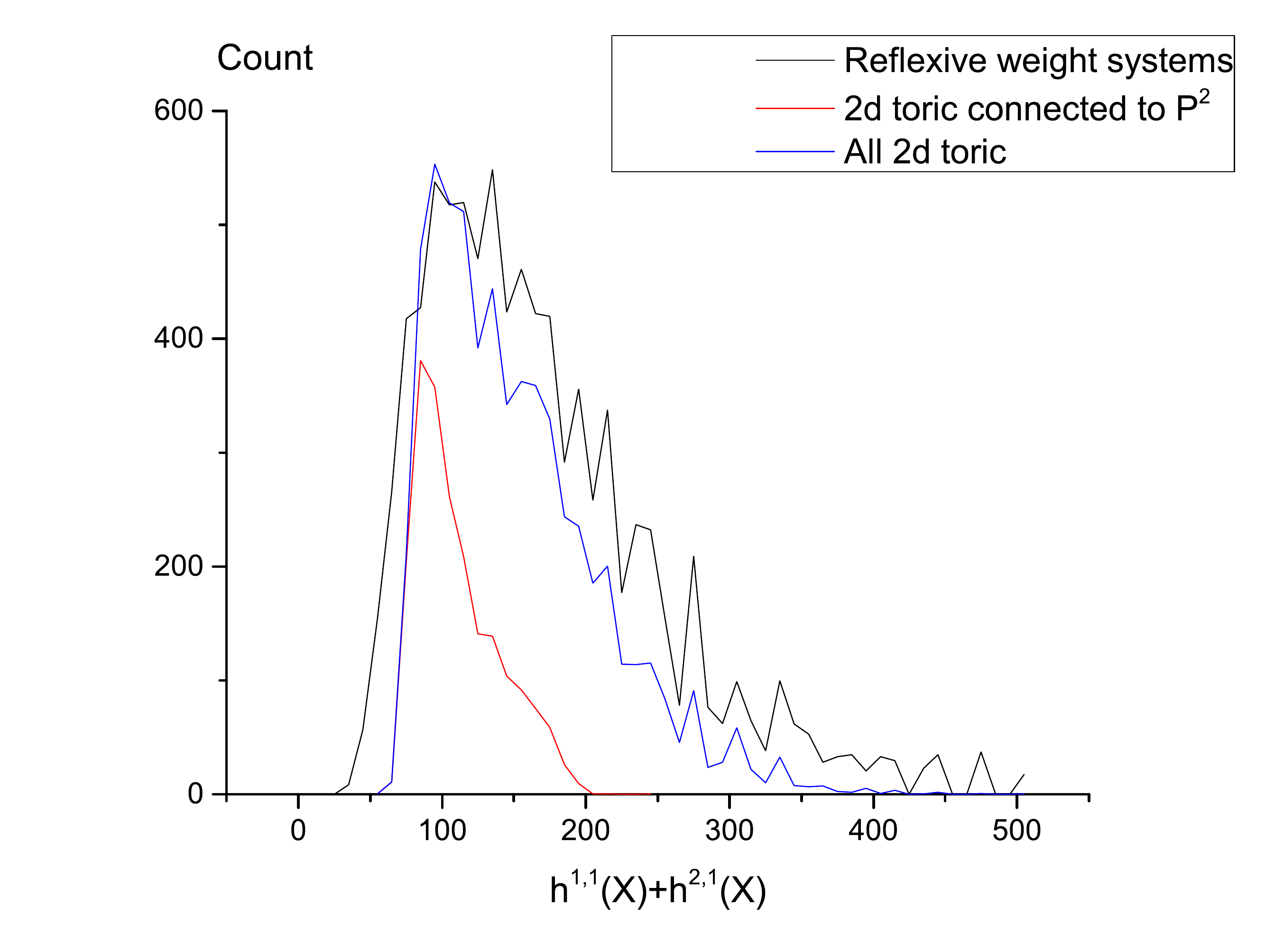}
\end{center}
\caption[x]{\footnotesize  Comparison of the distribution of Hodge
  numbers in the known Kreuzer-Skarke Calabi-Yau threefold database
  with the set of Hodge numbers for generic elliptically fibered CY
  threefolds over both the full set of toric bases  \cite{mt-toric}
and the subset in the connected set ${\cal C}_3$ related to $\P^2$ by
blow-ups and blow-downs that do not introduce $(4, 6)$ points. 
In order to match up with Figure~\ref{f:compare-ks}, we only choose
the subset in the Kreuzer-Skarke database that corresponds to
reflexive weight systems \cite{Skarke:1996hq}.
}
\label{f:3D-comparison}
\end{figure}

Though the
fourfolds we encounter in the Monte Carlo exploration have relatively
small Hodge numbers compared to the limits realized in  the full set
of known Calabi-Yau fourfolds, the Hodge numbers are
clustered in a region that is not far from the peak of the
distribution in the set of known fourfolds, at least for those that arise from reflexive
transverse weight systems. 
Figure~\ref{f:compare-ks} compares the distribution of $h^{1,1}+h^{3,
  1}$ encountered in the Monte Carlo runs to the distribution found
for a particular set of fourfold constructions, namely the transverse weight
systems found in \cite{lsw-4d} that give reflexive polytopes.
Note that the Monte Carlo distribution is much more peaked than that
for the known fourfolds.  There are several possible reasons for this.
First, bases with small $h^{1, 1}(B)$ can support a wide range of
``tunings'' of the Weierstrass model corresponding to distinct
codimension one and two singularity structures giving distinct
Calabi-Yau fourfolds with relatively small Hodge numbers over the same
base, while bases giving elliptic Calabi-Yau's with larger Hodge
numbers admit fewer tunings (see {\it e.g.} \cite{Johnson-WT}).  This
in general increases the number of fourfolds at small Hodge numbers
disproportionately to the number of bases.  The fourfolds that do not
admit elliptic fibrations with section may also be more common at
smaller Hodge numbers.  At larger Hodge numbers, there are fourfolds
that support non-Higgsable $E_8$ factors that arise from toric
constructions with $(4, 6)$ curves, which are not included in this
analysis.  These may increase the number of fourfolds with higher
Hodge numbers relative to the distribution of bases found in the Monte
Carlo.  Finally, there can be many weight systems that give rise to
the same Calabi-Yau, which may artificially enhance the distribution
of weight systems in certain regimes.  These issues make the
comparison in Figure~\ref{f:compare-ks} a rather rough analogy, but
the rough agreement between the regions of the peak suggests that with
the preceding caveats, somewhat similar distributions may be sampled
by the two different approaches.  To get a sense of the significance
of this comparison, we have considered a similar analysis in the case
of elliptic threefolds, with results shown in
Figure~\ref{f:3D-comparison}. In that graph we compare the set of
Hodge numbers for an analogous class of weight systems that give known
Calabi-Yau threefolds from hypersurfaces in toric varieties
\cite{Skarke:1996hq, akms, Kreuzer-Skarke} (data available at
\cite{Kreuzer-Skarke-data}) to generic elliptic fibrations over the
full set of toric base surfaces that support elliptic Calabi-Yau
threefolds (identified in \cite{mt-toric}) and the subset in the
connected set ${\cal C}_3$ related to $\P^2$ by blow-ups and
blow-downs that do not introduce $(4, 6)$ points.  In the 3D case we
see that the connected set has a similar shape but somewhat smaller
size and lower Hodge numbers from the complete set of toric bases.
And we see a similar rough agreement between the peaks of the
distributions, which are all well below the largest possible Hodge
numbers realized for threefolds.  As in the fourfold case, the
distribution from the connected set of base surfaces is more peaked
and undercounts the number of threefolds at both small and large Hodge
numbers.  In the case of threefolds explicit consideration of the
distributions shows that the reasons given above seem to characterize
the differences between the distributions accurately.  Note in
particular that there are many distinct weight systems that can
characterize the same reflexive polytope and elliptic threefold; this
explains the excess in the graph of the distribution of weight systems
compared to toric bases at large Hodge numbers seen in
Figure~\ref{f:3D-comparison}.  In fact, at very large Hodge numbers
all known threefolds are elliptic fibrations over toric bases, with
little tuning possible.

Analyses from several points of view
\cite{akms, Hodge,  Candelas-cs,
Gray-hl, Johnson-WT, Anderson-aggl} suggest that
in fact most or all known
Calabi-Yau threefolds and fourfolds with large Hodge numbers admit an
elliptic fibration.  Thus, both for threefolds and for fourfolds
we may expect that a complete  analysis of the bases involved, including
tunings of the generic Weierstrass model, may give a good
picture of the set of possible elliptic Calabi-Yau manifolds.
In any case, the rough similarity between Figure~\ref{f:compare-ks} and the
fairly parallel 3d analysis 
depicted in Figure~\ref{f:3D-comparison}
suggests that the Monte
Carlo analysis of threefold bases is exploring a reasonably
representative sample of the bases associated with a significant part of the
space of known fourfolds.  As in the case of threefold bases, we expect that our Monte
Carlo is missing an even larger number of bases that have $(4, 6)$
curves, associated with non-toric threefolds that support Calabi-Yau
fourfolds giving F-theory models with $E_8$ gauge factors.  It would
be nice to extend the kind of analysis we do in this paper to include
these other bases, though this is technically more complicated than in
the simpler case of base surfaces for elliptic threefolds.

\subsubsection{The number of threefold bases}
\label{sec:number}

The number of bases in the connected set we are exploring appears to
be quite large.  In particular, the tendency of the Monte Carlo runs
to enter separated domains in the graph of connected bases indicates
that the total number of bases available is at least
much larger than $10^5$.
The individual Monte Carlo runs we have carried out generally do not
hit bases with $h^{1, 1}(B) \lesssim 50$ once they have
  ``thermalized'' after 1000 or more steps.
To get a normalization on the distribution and thus estimate the total
number of bases, we have carried out a sequence of runs in which we
have placed an artificial upper bound on the Picard number of the
base.  In particular, we have done 10 Monte Carlo runs of
30,000 steps each with upper bounds $h^{1, 1}(B) \leq 5k+2$ for 
each
$k = 1,
\ldots, 13$. We again
ignore the first 1000 bases in all the statistical analyses. Using the appropriate weighting factors,
this gives an estimate of the distribution of bases in each bounded
range of $h^{1, 1}(B)$.

\begin{figure}
\begin{center}
\includegraphics[height=8cm]{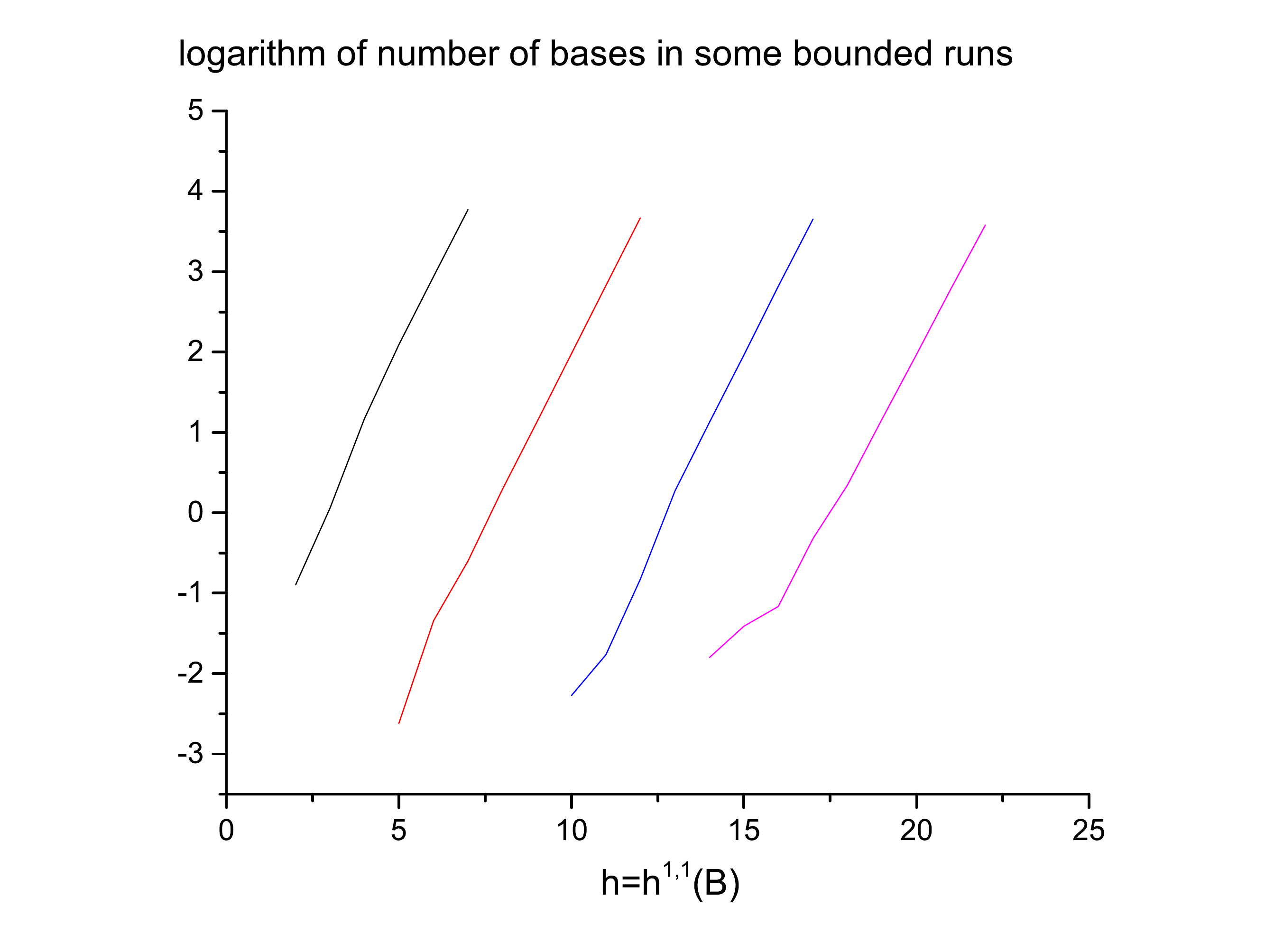}
\end{center}
\caption[x]{\footnotesize  Distributions of bases in connected sets
  with an upper bound $h^{1, 1}(B) \leq 5k+2$ for $k = 1, 2, 3,
  4$. Only the relative numbers are plotted, so that those curves can be
  displaced vertically and connected.}
\label{f:bounded-distributions}
\end{figure}

The  (logarithmically scaled) distributions of bases for the first few
values of $k$ are shown in Figure~\ref{f:bounded-distributions}. 

To estimate the total number of bases
in ${\cal C}$ we can combine the distributions
from the bounded runs.  We define
\begin{equation}
N  (h) = |\{B \in{\cal C}:h^{1, 1} (B) = h\} | \,.
\end{equation}
We know that $N (1) = 1$ (from $B =\P^3$), and it is not hard
to determine that 
$N (2) = 27$ (from
$\P^1 \times \P^2$, 12 distinct
nontrivial $\P^1$ bundles over $\P^2$ and 
14 distinct nontrivial
$\P^2$ bundles
over $\P^1$; there is also one toric base with $h^{1, 1}(B) = 2$
and an $E_8$
divisor
 --- the $\P^1$ bundle over $\P^2$ with twist 18
--- that is not in the connected graph ${\cal C}$.)
As a check on our methodology, the ratio $N (2)/N (1) = 27$ is
correctly reproduced to good accuracy by Monte Carlo runs with a low
bound on $h$.

We denote the number of bases with $h^{1,1}(B)=h$ 
encountered
in the experiment $h^{1,1}(B)\leq m$ by ${\cal N}_m(h)$. 
The numbers are geometrically averaged among multiple runs.
Then the run at $k = 1$ gives an estimate of
$N (7) $, using the experimental ratio ${\cal N}_7(7)/{\cal N}_7(2)$
and the fact that $N(2)=27$:
\be
N(7)\cong 27\cdot\frac{{\cal N}_7(7)}{{\cal N}_7(2)}.
\ee
From the run at $k = 2$ we can
 use the experimental value ${\cal N}_{12} (12)/{\cal N}_{12} (7)$ to estimate $N(12)$.  Repeating
this process we can give a rough estimate for
\begin{equation}
N (h) \cong 27 \times \frac{{\cal N}_7 (7)}{{\cal N}_7 (2)} 
\times \frac{{\cal N}_{12} (12)}{{\cal N}_{12} (7)}  \times \cdots
\times \frac{{\cal N}_{h'} (h')}{{\cal N}_{h'} (h' -5)} 
\times \frac{{\cal N}_{h'+5} (h)}{{\cal N}_{h'+5} (h')}  \,,
\end{equation}
where $ h' \equiv 2$ (mod 5) and $h-5 < h' \leq h$. Finally when
  $h\geq 67$, the proportion of bases
at each $h$ is significant enough that we can employ the data from the 100 unbounded runs, and $N(h)$ can be estimated by
\be
N(h)\cong N(67)\cdot\frac{{\cal N}_{unbounded}(h)}{{\cal N}_{unbounded}(67)}.
\ee

\begin{figure}
\begin{center}
\includegraphics[height=8cm]{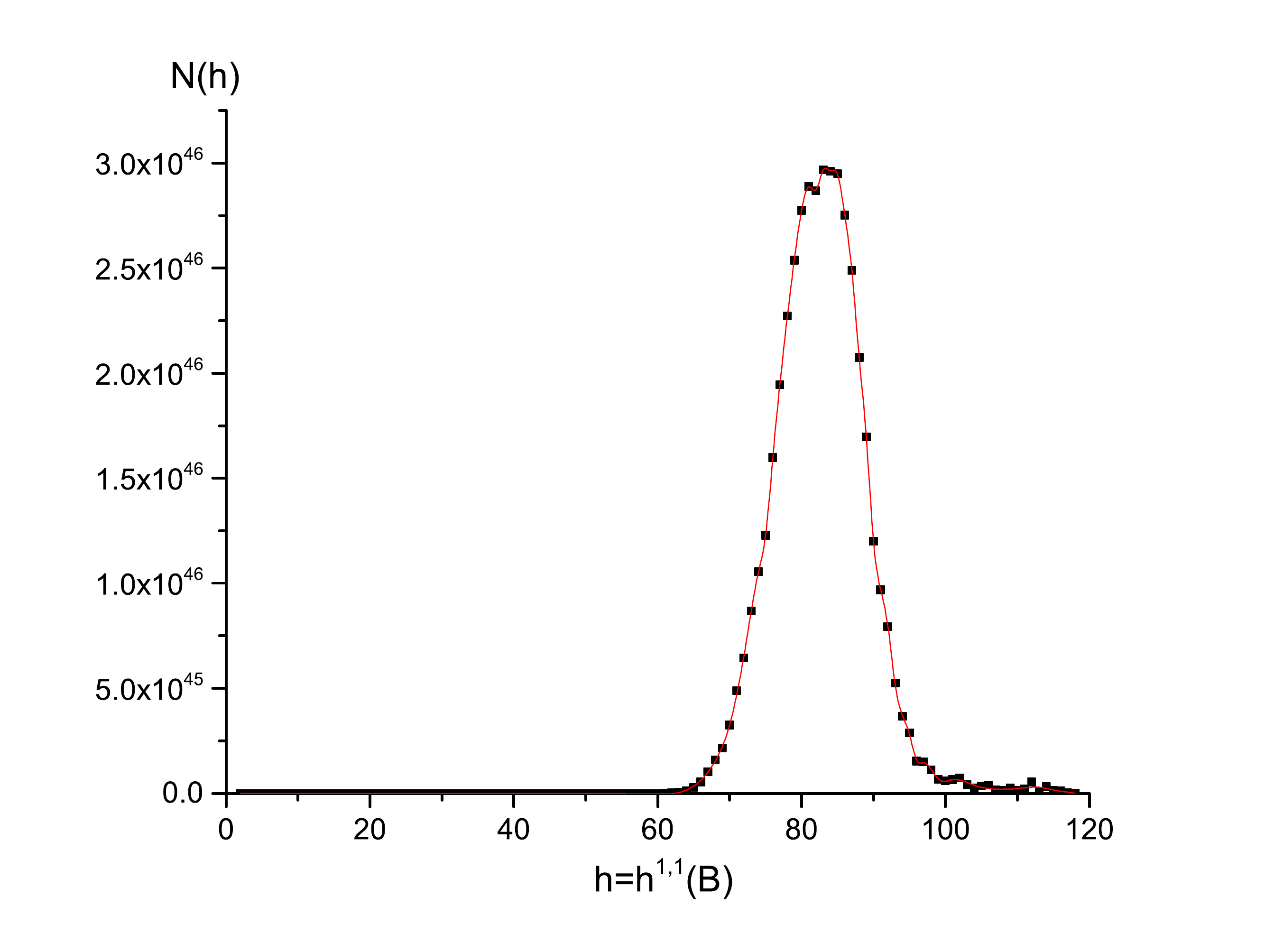}
\end{center}
\caption[x]{\footnotesize  Rough estimation of the number of bases $B
  \in{\cal C}$ with $h^{1, 1}(B) = h$.}
\label{f:numbers-graph}
\end{figure}

\begin{figure}
\begin{center}
\includegraphics[height=8cm]{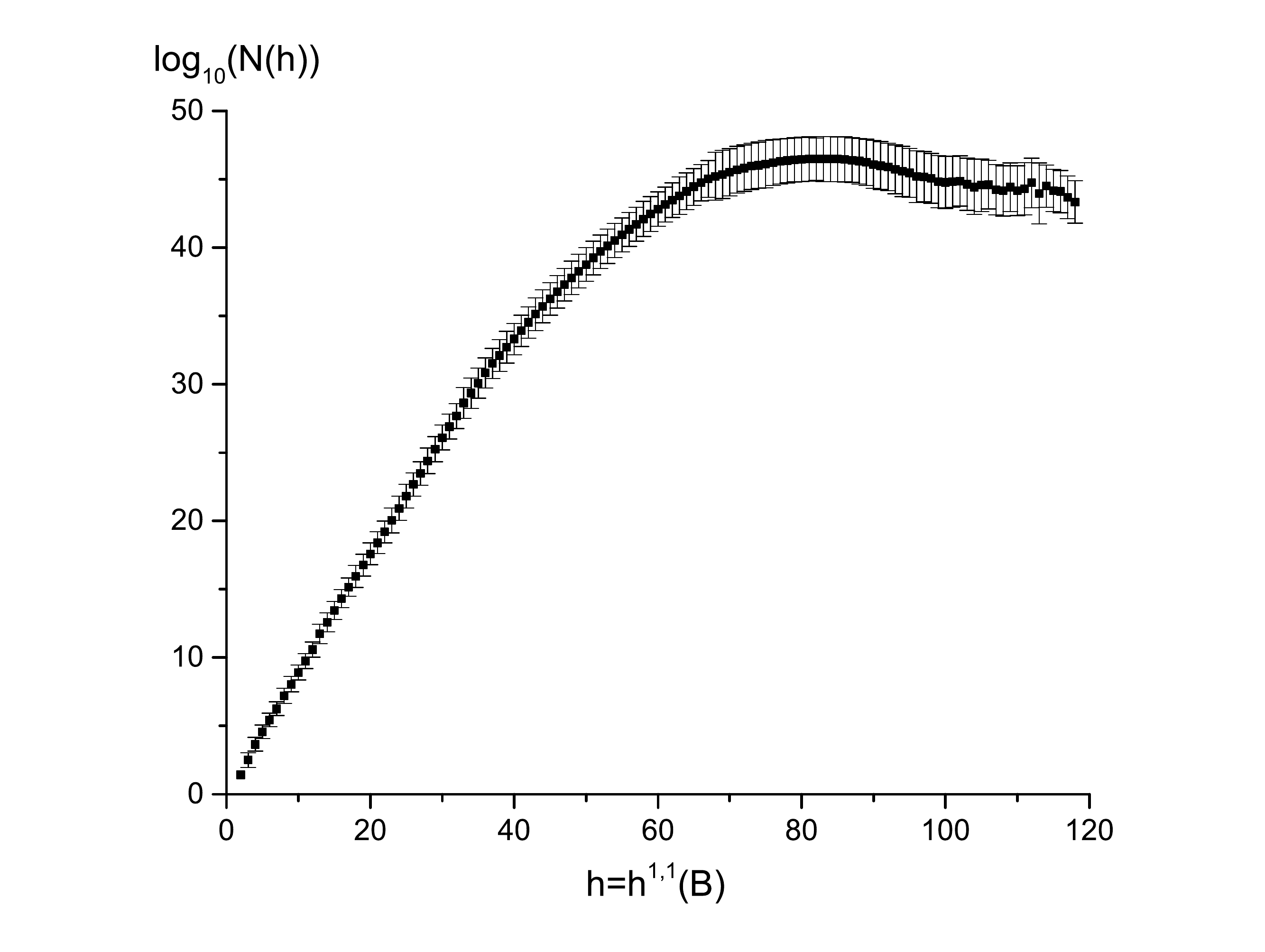}
\end{center}
\caption[x]{\footnotesize  Rough estimation of $\log_{10}(N(h))$ with error bars, where $N(h)$ is the number of bases $B
  \in{\cal C}$ with $h^{1, 1}(B) = h$.}
\label{f:lognumbers}
\end{figure}

The resulting estimations of $N (h)$ are graphed in
Figure~\ref{f:numbers-graph}. We also plot $\log_{10}(N(h))$ in
Figure~\ref{f:lognumbers}, with the standard deviation. It turns out
that in the region $h\leq 35$, the number of bases grows
exponentially. In the region $35 \leq h\leq 60$, the exponential
growth slows down. Finally the number of bases reaches a peak at
$h\cong 82$.

Summing these approximate values, we have a very rough estimate
\begin{equation}
| {\cal C}|= \sum_{h = 1}^{\infty} {\cal N} (h)
\sim { 10^{48\pm 2}} \,.
\end{equation}

\subsubsection{Number of possible flops on each base}

One possible explanation for the exponential growth of the number of bases
with $h^{1,1}(B)$ is the existence of many
flops (see Figure~\ref{f:flop})
on typical bases with large
$h^{1,1}(B)$. Since a flop does not change the
rays or the set of monomials, the geometric
non-Higgsable gauge group content
of the associated F-theory compactification does not
change under a flop. 
Because $v_i+v_j=v_k+v_l$, $f$ and $g$ also
vanish to the same order
on the toric curves $v_i v_j$ and $v_k v_l$. Hence if $B_1\in{\cal C}$,
and $B_2$ can be related to $B_1$ by a flop, then $B_2\in{\cal C}$
always holds.
Note, however, that because the curves lie on different combinations
of divisors after a flop, the matter content of an F-theory model,
associated with the codimension two singularities on curves that live
in each divisor, may
change under a flop.

\begin{figure}
\begin{center}
\includegraphics[width=12cm]{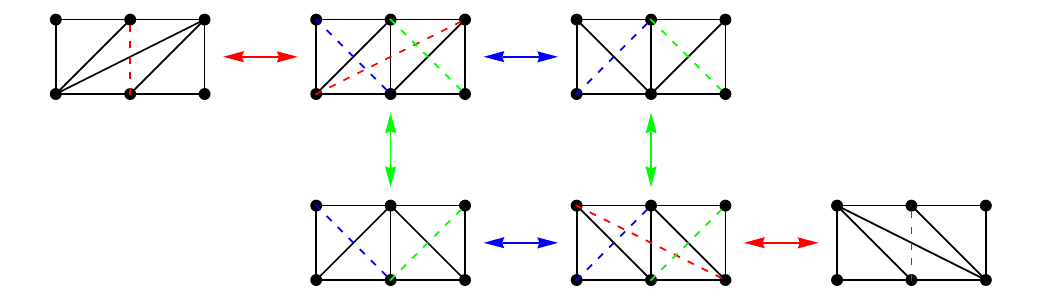}
\end{center}
\caption[x]{\footnotesize  Different cone configurations connecting
  six rays associated with coplanar vertices in the lattice $N =\Z^3$
  are connected by flops.  In two of these configurations there is
  only one possible flop, in another two there are two possible flops,
and in the other two there are three possible flops.}
\label{f:flops}
\end{figure}

If there are $n$ possible flops on a base $B\in{\cal C}$, and each
flop is isolated, then there are $2^n$ bases in ${\cal C}$ that can be
related to $B$ by a sequence of flops.  A flop can occur when 4
vertices associated with rays $v_i$ are coplanar and connected by 2d
cones as in Figure~\ref{f:flop}.  When more than 4 vertices are
coplanar and connected, multiple connected flops can be possible.  In
this case, the total number of distinct bases in the set connected by
flops may not be exactly $2^n$.  Performing a flop may, for example,
destroy the possibility of others among the $n$ possible flops.  Or it
may bring in new possible flops, so the $2^n$ is just an estimation.
An example of 6 vertices, with six distinct triangulations connected
by flops, is shown in Figure~\ref{f:flops}.  In different subsets of these
configurations the number of possible flops is  $n =$ 1, 2, or 3, showing
that $2^n$ can over or underestimate the actual number of
configurations connected through flop sequences.
  All of those bases give rise to the
same gauge groups and similar physics in F-theory compactifications,
though precise details including the (geometric) matter spectrum may
differ between the constructions.  Despite the differences caused by
connected sets of flops, $2^n$ can be viewed as a good order of
magnitude estimate for the number of distinct bases associated with
different triangulations of a single structure connected by flops.

\begin{figure}
\begin{center}
\includegraphics[width=10cm]{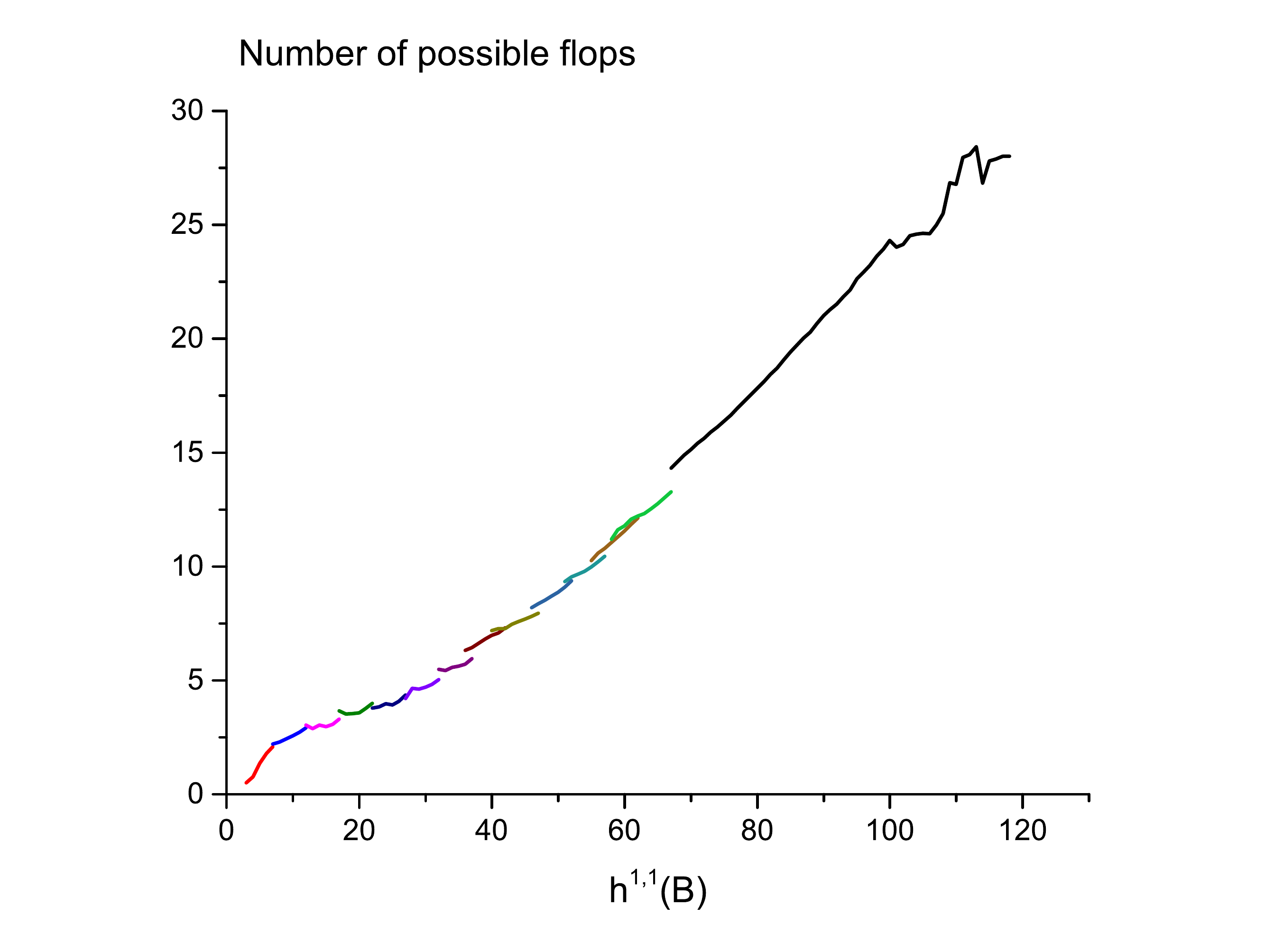}
\end{center}
\caption[x]{\footnotesize Average number of possible flops on $B$ as a
  function of $h^{1,1} (B)$. For  $h^{1,1}(B)>67$, the numbers are
computed from the 100 unbounded runs. For the lower values of
$h^{1,1}(B)$, they are computed from the bounded runs.}
\label{f:flopcount}
\end{figure}

The average numbers of possible flops on bases with different
$h^{1,1}(B)$ are plotted in Figure~\ref{f:flopcount}.  The number of
flops grows almost linearly from $h^{1,1}(B)=50$ to $h^{1,1}(B)=100$.
Comparing to Figure~\ref{f:lognumbers}, we can see that even if we
divide the total number of bases for a given $h^{1,1}(B)$ by $2^n$,
where $n$ is the average number of possible flops, the number of
distinct triangulation types of bases still grows exponentially. Hence
there are still approximately $10^{48}/2^{20} \sim 10^{42}$ distinct
bases that cannot be related by flops.

\subsection{Distribution of non-Higgsable group factors}

The geometrically non-Higgsable gauge groups and matter that  arise on
divisors and curves in the F-theory base $B$ provide a convenient
structure with which to characterize both the geometry of elliptic
Calabi-Yau manifolds and the physics of the associated F-theory
compactifications \cite{clusters, Hodge,
Anderson-Taylor, ghst, 4d-NHC, Halverson-WT}.  While a full analysis of the
physics of a given F-theory model would involve many additional
considerations, including tuning of enhanced gauge symmetries or
matter fields, G-flux, brane world-volume fields, 
{\it etc.}, the
non-Higgsable geometry provides a starting point for such analysis.
In this section 
and the following section
we look at generic features of non-Higgsable gauge
groups that arise on the bases found in our Monte Carlo study.  In
\S\ref{sec:matter}, we look at codimension two singularities
associated with matter fields and other structure.

\subsubsection{Number of factors in $G$}

Essentially all the bases found in the Monte Carlo runs had some divisors
supporting non-Higgsable gauge factors.  The only exceptions were in
the first few bases encountered in each run, before
``thermalization''.

\begin{figure}
\begin{center}
\includegraphics[width=10cm]{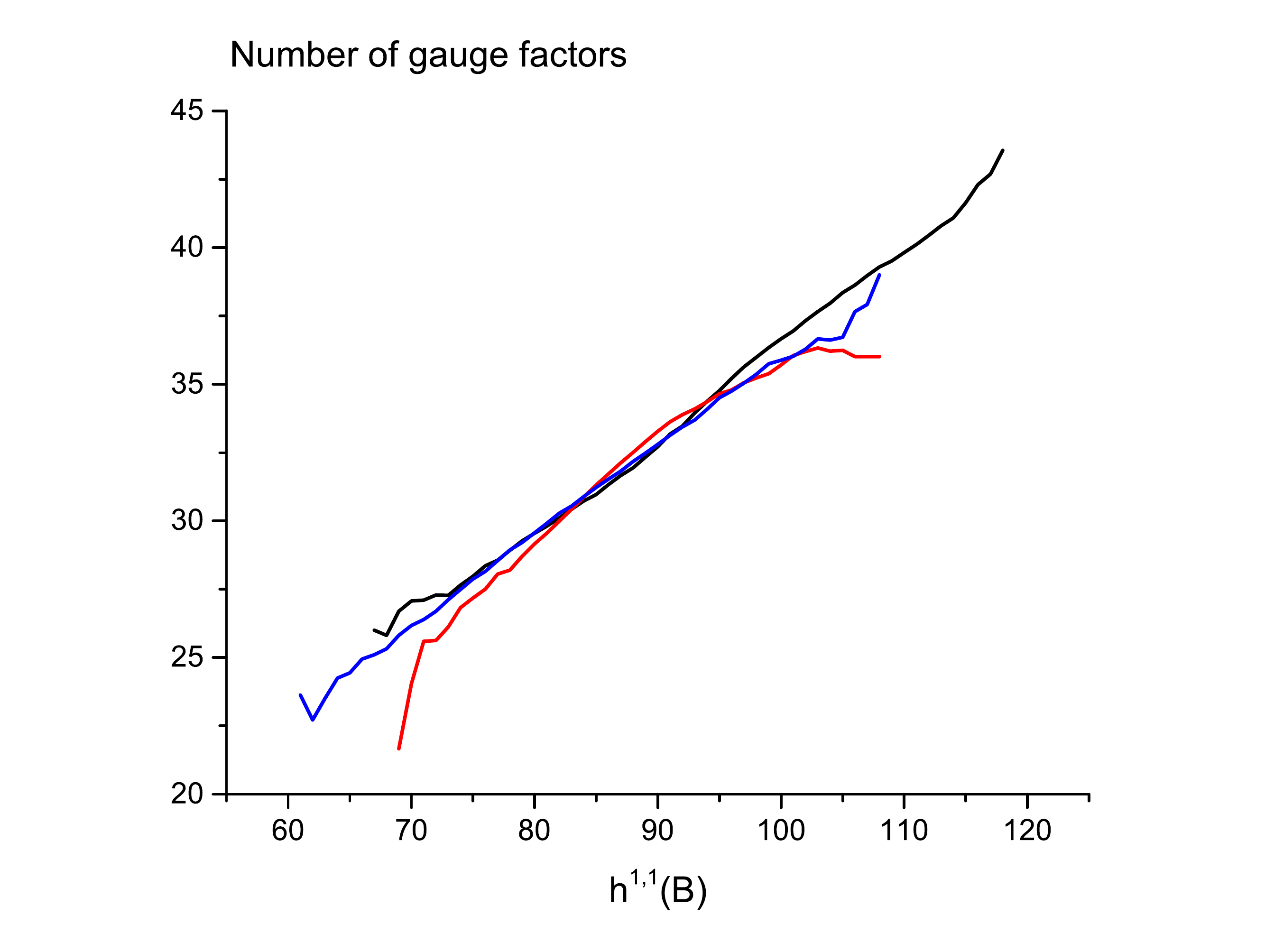}
\end{center}
\caption[x]{\footnotesize  Average number
 of factors in $G$ as a function of $h^{1,1} (B)$, in three different unbounded runs with 100,000 samples. }
\label{f:g-factors-3}
\end{figure}

\begin{figure}
\begin{center}
\includegraphics[width=10cm]{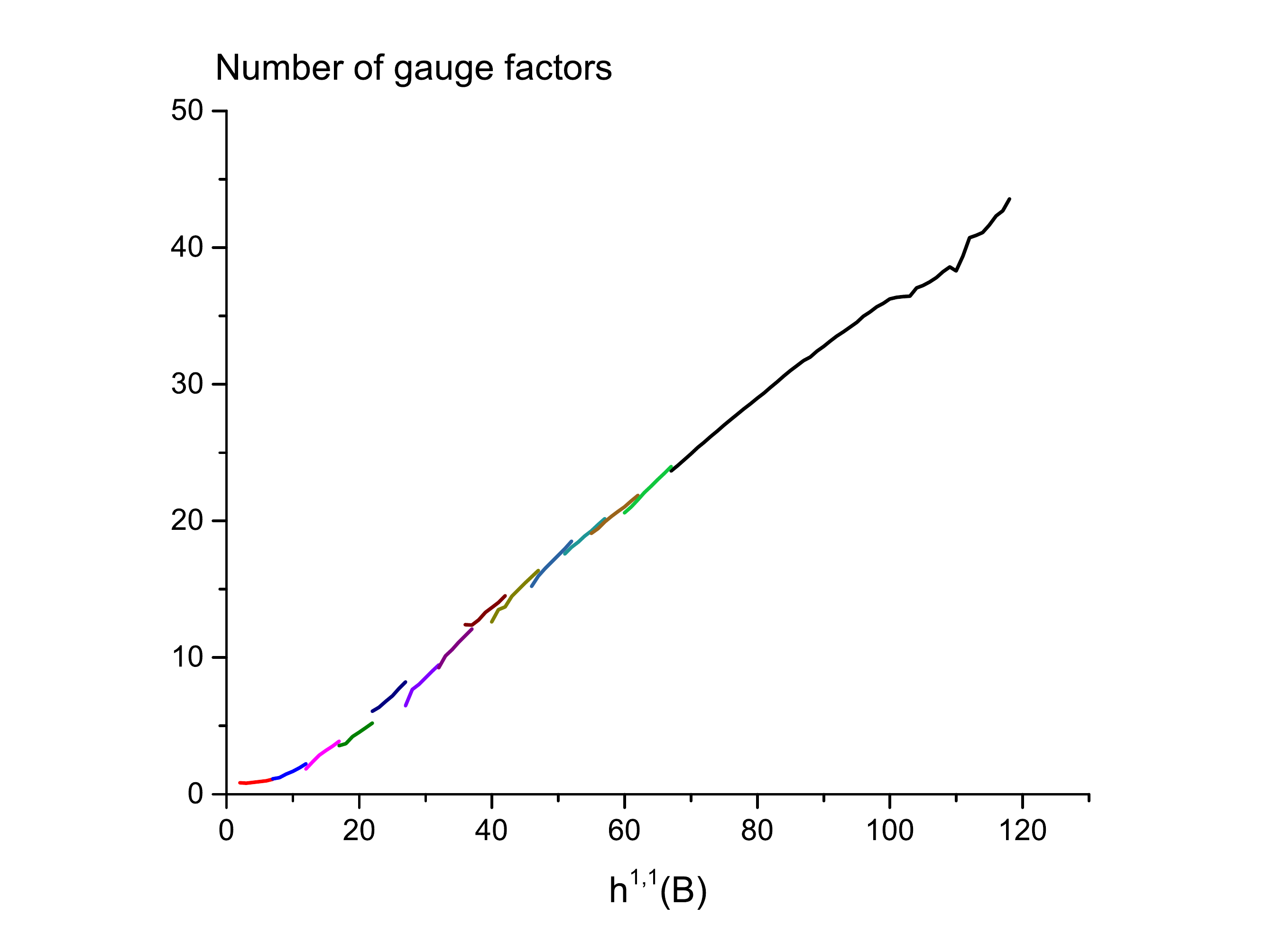}
\end{center}
\caption[x]{\footnotesize Average number of factors in $G$ as a
  function of $h^{1,1} (B)$. For the part of the graph with
  $h^{1,1}(B)>67$,  averages are
computed from the 100 unbounded runs. For the lower values of
$h^{1,1}(B)$, averages are computed from the bounded runs.}
\label{f:g-factors-average}
\end{figure}

The number of factors in $G$ grows roughly linearly with $h^{1,1}(B)$.
This is shown in Figure~\ref{f:g-factors-3}
for several different individual runs, and averaged over runs in
Figure~\ref{f:g-factors-average}.
For bases with $h^{1, 1}(B)$ between  40 and 100, the fraction of
divisors on any base that support a non-Higgsable gauge factor is
roughly 35--40\%.  The fraction is slightly smaller for low $h^{1,
  1}(B)$, which is not surprising as the divisors can more easily have
positive normal bundles when there are fewer rays in the toric fan,
and non-Higgsable gauge factors are associated with negative
contributions to the normal bundle \cite{4d-NHC}.
Note that this fraction of divisors supporting non-Higgsable gauge
factors is significantly smaller than in the case of $\P^1$-bundle
bases studied in \cite{Halverson-WT} (see Figure 12 in that paper).
This can be understood because of the special geometric structure of
the $\P^1$-bundle bases, which is dominated by an essentially 2d
structure of the base surface $S$ for the $\P^1$ bundle, so that the
statistics there are closer to what is expected from non-Higgsable
groups on toric base surfaces \cite{mt-toric}, where the divisors form
a linear chain, with one- or two-factor non-Higgsable clusters
separated by $-1$ curves on surfaces of large $h^{1, 1}(S)$.  The
smaller fraction of divisors we find in the Monte Carlo analysis is
presumably a manifestation of the truly 3D nature of the toric
threefold bases.  An interesting question is whether this fraction will
be similar for non-toric bases, which do not have the intersection
structure on divisors associated with the triangulation of an $S^2$ by
the divisor rays as in the toric threefold case.

\subsubsection{Distribution of gauge factors}

\begin{table}
\begin{center}
\begin{tabular}{|c|c|c|c|c|}
\hline
\hline
SU(2)$_{III}$&SU(2)$_{IV}$&SU(2)&SU(3)&$G_2$\\
\hline
$6.1\pm 1.7$ & $7.5\pm 1.5$ & $13.6\pm 1.6$ & $2.0\pm 0.6$ & $9.7\pm 1.8$\\ 
\hline
\hline
SO(7)&SO(8)&$F_4$&$E_6$&$E_7$\\
\hline
 $4\times 10^{-6}\pm 2\times 10^{-5}$ 
&$1.0\pm 0.6$ & $2.8\pm 1.1$ & $0.3\pm 0.4$ & $0.2\pm 0.5$\\
\hline
\hline
\end{tabular}
\end{center}
\caption[x]{\footnotesize Average number of times each 
non-Higgsable
gauge group
  factor appears on a base, with standard deviation computed among the
  100 runs.} 
\label{t:gaugegroup}
\end{table}

\begin{figure}
\begin{center}
\includegraphics[width=13cm]{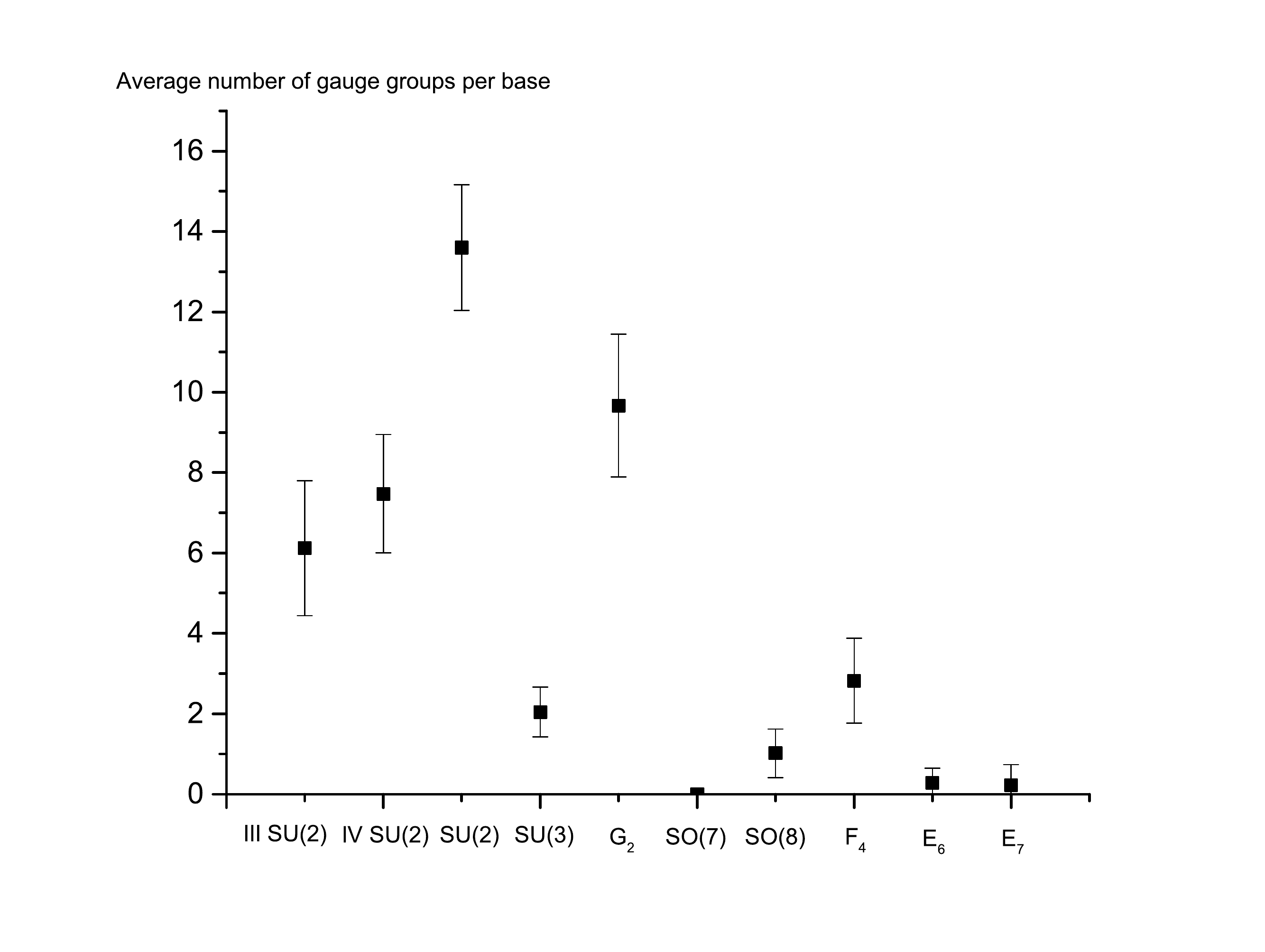}
\end{center}
\caption[x]{\footnotesize  Average number
 of times each gauge group factor appears per base, with standard
 deviation computed among the 100 runs. } 
\label{f:gaugegroup}
\end{figure}

\begin{figure}
\begin{center}
\includegraphics[width=13cm]{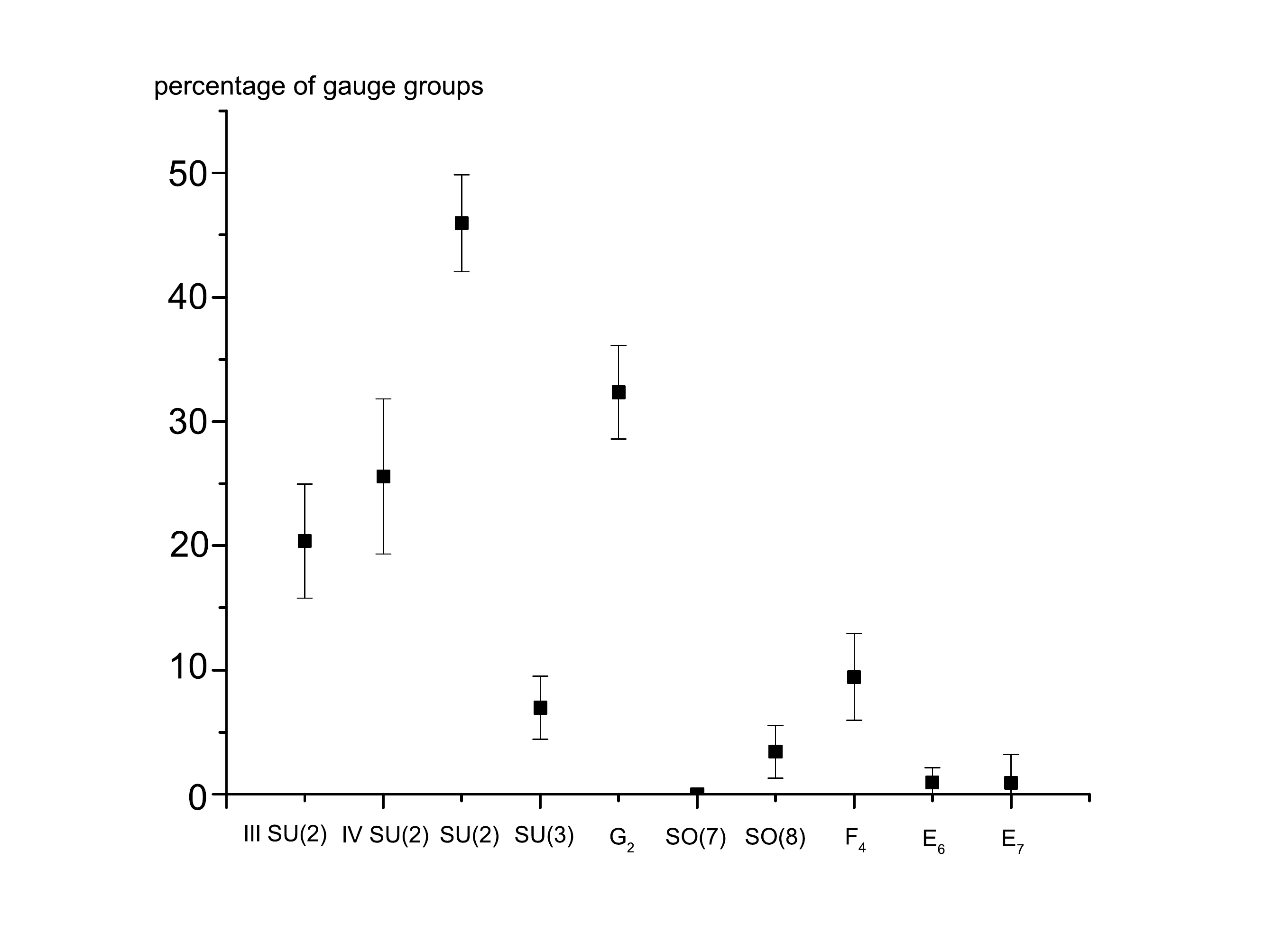}
\end{center}
\caption[x]{\footnotesize Average percentage appearance of each gauge
  group factor, with standard deviation computed among the 100 runs. }
\label{f:gaugegroupfrac}
\end{figure}

We have listed the average numbers of times that each individual
non-Higgsable gauge group factor arises on a typical base in
Table~\ref{t:gaugegroup} and Figure~\ref{f:gaugegroup}.  The
percentage of the time that each gauge factor arises among all the
gauge group factors is listed in Table~\ref{t:gaugegroupfrac} and
Figure~\ref{f:gaugegroupfrac}.  We also list the percentage of bases
with a specific gauge group factor in Table~\ref{t:gaugegroupbw}.

\begin{table}
\begin{center}
\begin{tabular}{|c|c|c|c|c|}
\hline
\hline
SU(2)$_{III}$&SU(2)$_{IV}$&SU(2)&SU(3)&$G_2$\\
\hline
$20\pm 5$ & $26\pm 6$ & $46\pm 4$ & $7\pm 3$ & $32\pm 4$ \\
\hline
\hline
SO(7)&SO(8)&$F_4$&$E_6$&$E_7$\\
\hline
 $1\times 10^{-5}\pm 7\times 10^{-5}$ 
&$3.4\pm 2.1$ & $9\pm 3$ & $0.9\pm 1.1$ & $0.9\pm 2.3$\\
\hline
\hline
\end{tabular}
\end{center}
\caption[x]{\footnotesize Average percentage of each gauge group factor, with standard deviation computed among the 100 runs.}
\label{t:gaugegroupfrac}
\end{table}

\begin{table}
\begin{center}
\begin{tabular}{|c|c|c|c|c|c|c|c|}
\hline
\hline
SU(2)&SU(3)&$G_2$&SO(7)&SO(8)&$F_4$&$E_6$&$E_7$\\
\hline
$99.999\pm 0.001$ & $83\pm 11$ & $99.93\pm 0.07$ & $0.0004\pm 0.002$ & $59\pm 21$ & $94\pm 21$ 
&$26\pm 31$ & $18\pm 37$\\
\hline
\hline
\end{tabular}
\end{center}
\caption[x]{\footnotesize Average percentage of bases with a specific
  gauge group factor, with standard deviation computed among the 100 runs.}
\label{t:gaugegroupbw}
\end{table}

It turns out that the gauge factors SU(2) and $G_2$ are mostly
dominant. The gauge factors $F_4$ and SU(3) also generally arise on a
typical base, with an average number of appearances higher than 1 in
each case. For the other gauge group factors, their appearance seems
to characterize some ``local feature'' of the part of landscape
covered by a particular run. The gauge group SO(7) is the most rare
one; from these statistics, on a typical base one does not expect the
appearance of SO(7).

Comparing to the distribution of gauge factors found on $\P^1$-bundle
bases \cite{Halverson-WT}, the percentage of SU(3), SO(8) and $E_6$
gauge groups are much higher than the corresponding total percentage
values in Table.\ 3 of \cite{Halverson-WT}.  But the percentage of
SO(7) and $E_7$ gauge factors found here are much lower.  Because we do not move
across regions with codimension-2 (4,6) singularities, it is natural
to expect that gauge groups with high rank such as $E_7$ and $E_8$
will be much rarer in our Monte Carlo analysis. 
The relative frequency of $SO(7)$ factors in $\P^1$-bundle bases
likely comes from the basically 2d nature of the bases in that case.
$SU(2) \times SO(7) \times SU(2)$ is a standard non-Higgsable cluster
that arises on 2d base surfaces for elliptic Calabi-Yau threefolds
 that contain a chain of curves of self-intersections
$-2, -3, -2$ \cite{clusters}, and if such a sequence of curves appears
 in the base $S$ supporting the $\P^1$ bundle the same gauge group
 combination can appear as a non-Higgsable structure in
the resulting threefold base when the twist of the $\P^1$ bundle over
that non-Higgsable cluster is minimal.  The absence of these factors in our
Monte Carlo study  reflects the more intrinsically 3d structure
of the bases explored here. This explanation agrees with the
observation made in \cite{Halverson-WT}, that the percentages of
non-Higgsable
gauge
group factors that arise on the sections of the $\P^1$ bundle bases
(last line of Table 3 in that paper) are much higher for $SU(3)$,
$SO(8)$ and $E_6$ factors than on other divisors, which makes sense
since the sections are described geometrically by a broader class of
surfaces that locally corresponds more closely to the general set of
toric divisors in the toric threefold bases explored in the Monte
Carlo analysis here.
The percentages of gauge groups we find here indeed correspond
reasonably well with those found in \cite{Halverson-WT} when
restricted to sections of the $\P^1$-bundle base, suggesting that the
broad features of these results are fairly generic.  In particular,
the dominance of $G_2$ and $SU(2)$, the moderate level of appearance
of $SU(3), SO(8)$ and $E_6$, and the relative rarity of $SO(7)$ are
common features to these distributions.

\subsubsection{Distribution of gauge pairs}

As discussed in \cite{4d-NHC}, the only possible configurations of two
non-Higgsable gauge factors located on neighboring divisors are:
\be
SU(2)\times SU(2)\ ,\ SU(3)\times SU(2)\ ,\ SU(3)\times SU(3)\ ,\ G_2\times SU(2)\ ,\ SO(7)\times SU(2)
\ee
This follows from the requirement that there is not a (4,6)
singularity on the intersection of the two divisors, along with
monodromy conditions.
Such gauge pairs are naturally associated with codimension two
singularities supporting (geometric) matter that transforms as a
 field charged under both factors.

\begin{table}
\begin{center}
\begin{tabular}{|c|c|c|c|c|}
\hline
\hline
SU(2)$\times$SU(2)&SU(3)$\times$SU(2)&SU(3)$\times$SU(3)&$G_2\times$SU(2)&SO(7)$\times$SU(2)\\
\hline
$7.6\pm 1.9$ & $2.4\pm 0.9$ & $0.4\pm 0.4$ & $14\pm 3$ & $0\pm 0$\\
\hline
\hline
\end{tabular}
\end{center}
\caption[x]{\footnotesize Average number of appearances of
 each gauge pair on a base, with standard deviation computed among the 100 runs.}
\label{t:gaugepair}
\end{table}

\begin{figure}
\begin{center}
\includegraphics[width=13cm]{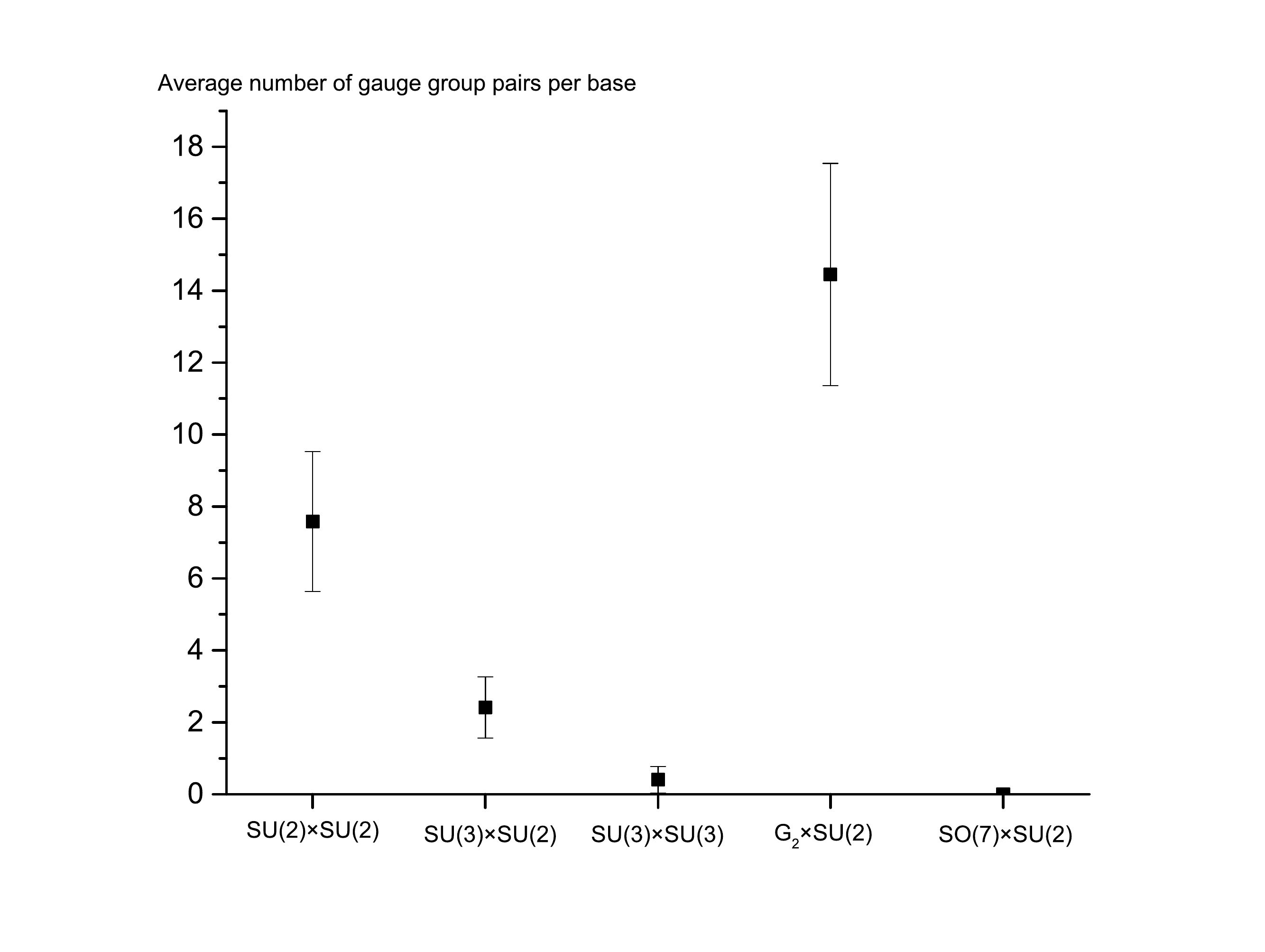}
\end{center}
\caption[x]{\footnotesize  Average number
 of appearances of
each gauge pair per base, with standard deviation computed among the 100 runs. }
\label{f:gaugepair}
\end{figure}

\begin{figure}
\begin{center}
\includegraphics[width=13cm]{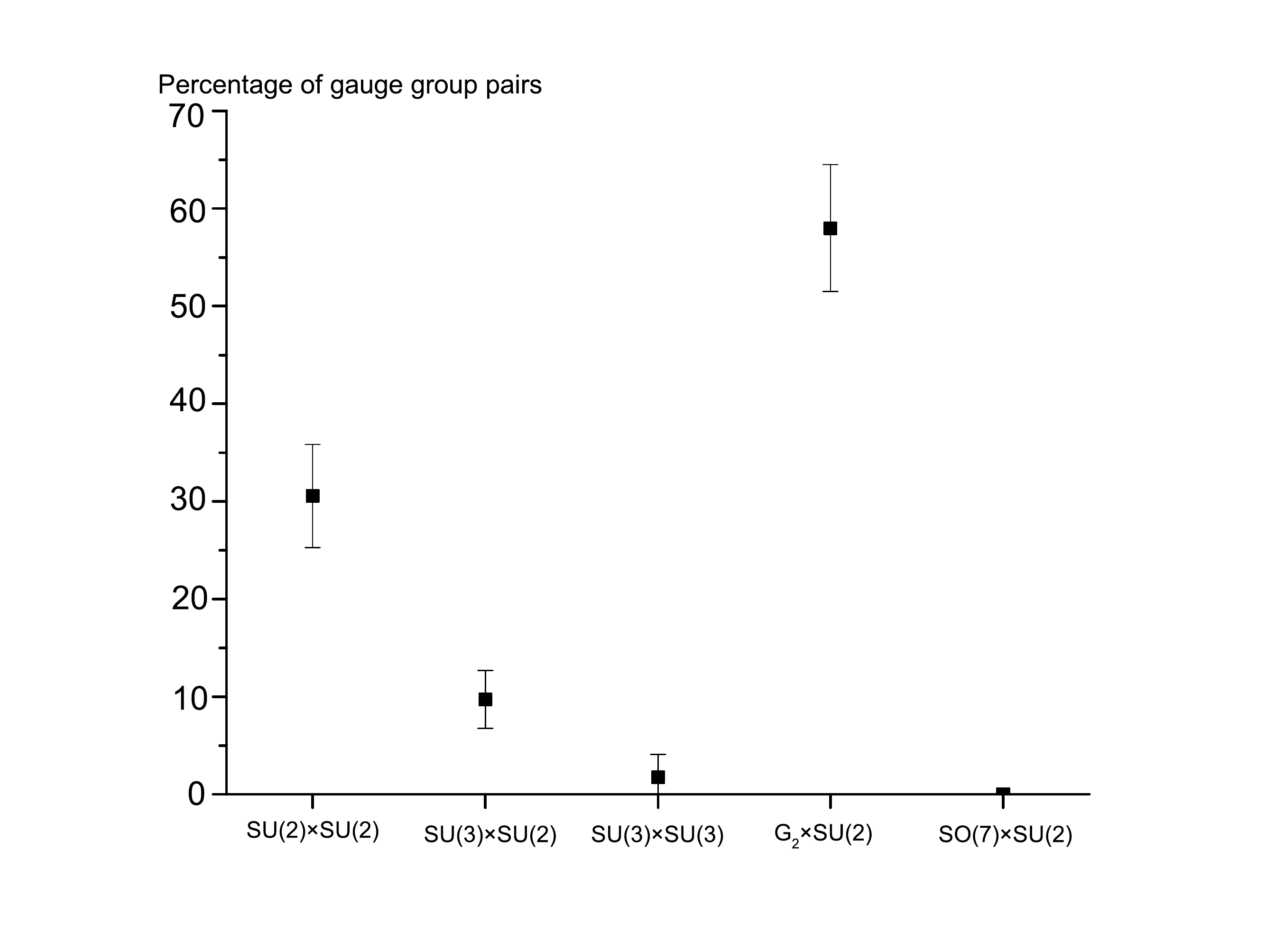}
\end{center}
\caption[x]{\footnotesize  Average percentage
 of each gauge pair, with standard deviation computed among the 100 runs. }
\label{f:gaugepairfrac}
\end{figure}

We have listed the average number of times each gauge pair
arises on a typical base
in Table~\ref{t:gaugepair} and Figure~\ref{f:gaugepair}.
The percentage of each gauge pair among all the gauge pairs is listed in Table~\ref{t:gaugepairfrac} and Figure~\ref{f:gaugepairfrac}.
We also list the percentage of bases with a specific gauge pair in
Table~\ref{t:gaugepairbw}.

\begin{table}
\begin{center}
\begin{tabular}{|c|c|c|c|c|}
\hline
\hline
SU(2)$\times$SU(2)&SU(3)$\times$SU(2)&SU(3)$\times$SU(3)&$G_2\times$SU(2)&SO(7)$\times$SU(2)\\
\hline
$31\pm 5$ & $10\pm 3$ & $1.7\pm 2.4$ & $58\pm 6$ & $0\pm 0$\\
\hline
\hline
\end{tabular}
\end{center}
\caption[x]{\footnotesize Average percentage of each gauge pair, with standard deviation computed among the 100 runs.}
\label{t:gaugepairfrac}
\end{table}

\begin{table}
\begin{center}
\begin{tabular}{|c|c|c|c|c|}
\hline
\hline
SU(2)$\times$SU(2)&SU(3)$\times$SU(2)&SU(3)$\times$SU(3)&$G_2\times$SU(2)&SO(7)$\times$SU(2)\\
\hline
$98\pm 6$ & $76\pm 14$ & $28\pm 16$ & $99.9\pm 0.7$ & $0\pm 0$\\
\hline
\hline
\end{tabular}
\end{center}
\caption[x]{\footnotesize Average percentage of bases with a specific
  gauge pair, with standard deviation computed among the 100 runs.}
\label{t:gaugepairbw}
\end{table}

The data on gauge pairs elaborates on
 the results presented in the
last section. The most dominant gauge pairs are SU(2)$\times$SU(2) and
$G_2\times$ SU(2);
SU(3) $\times$ SU(2) appears moderately frequently,
and the gauge pair SO(7)$\times$ SU(2) is so rare
that it never appears in our sampling runs. This indicates that the
average number of SO(7)$\times$SU(2) pairs
on a typical base could be lower
than $1\times 10^{-7}$.
The qualitative features of the distribution on pairs match with what
was found in \cite{Halverson-WT} for gauge pairs on divisors of which
one is a section of a $\P^1$-bundle base.

An interesting feature in the statistics is that for a typical base,
the gauge pair SU(3)$\times$SU(2) appears more than once, and more
than half of bases ($\sim 76$\%) support at least one
SU(3)$\times$SU(2) gauge pair.  Such a non-Higgsable gauge product
could act as the non-Abelian part of the standard model gauge group in
a MSSM-like scenario \cite{ghst}. We leave the detailed construction
of such phenomenological models to future work.

\subsection{Clusters}

\begin{figure}
\begin{center}
\includegraphics[width=10cm]{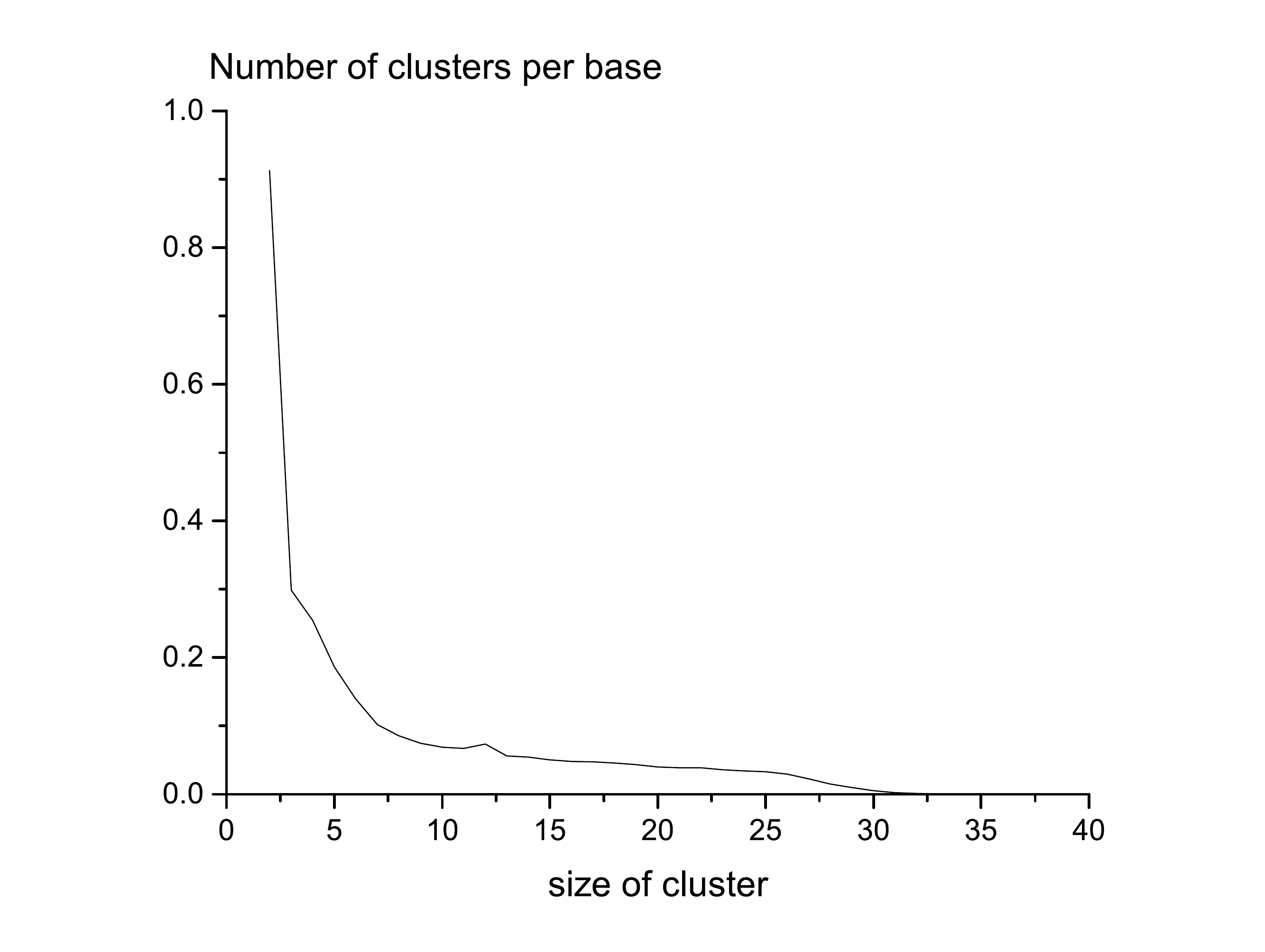}
\end{center}
\caption[x]{\footnotesize The average number of non-Higgsable clusters of each size on a base.}
\label{f:clustersize}
\end{figure}

As discussed in \cite{4d-NHC}, there are many possible non-Higgsable
clusters with size greater than two in 4d F-theory. Since the possible
configurations may be essentially arbitrarily complicated, bounded
only by the hypothetically finite number of threefold bases that
support elliptic Calabi-Yau fourfolds, it is not feasible to classify
all the large clusters. We only present some statistical data here.
On average, each base has a non-Higgsable gauge group with roughly 30
simple non-Abelian factors, of which $6.6\pm 1.6$ are single gauge
factors that are not contained in any larger non-Higgsable
clusters. Those gauge group components automatically include all the
SO(8), $F_4$, $E_6$ and $E_7$ gauge factors. 
On each base there are
$0.9\pm 0.5$ non-Higgsable clusters with size equal to two, and
$2.0\pm 0.5$ larger clusters. We plot the average numbers of non-Higgsable clusters of different size,
on a base in Figure~\ref{f:clustersize}.
The average cluster size is $3.3\pm
0.8$, including the single gauge groups. On each base we can find the
largest non-Higgsable cluster, and its average size is $16\pm 4$. All
the standard deviations are computed among the 100 runs.

\begin{figure}
\begin{center}
\includegraphics[width=10cm]{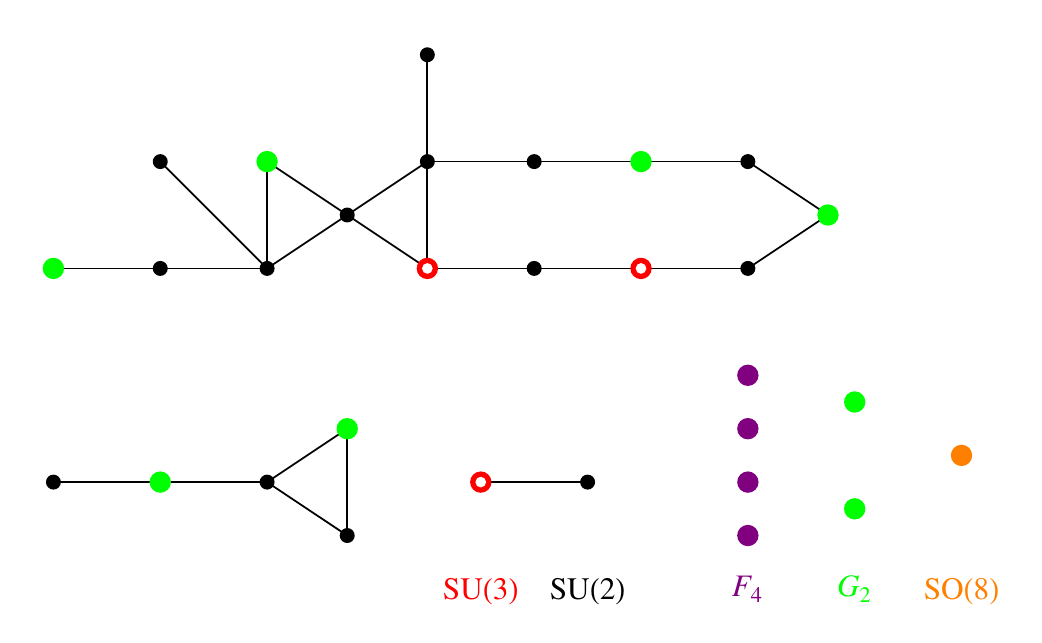}
\end{center}
\caption[x]{\footnotesize The non-Higgsable clusters of a fairly
typical base actually encountered
in the Monte Carlo study of toric threefold bases.  This base has
$h^{1, 1}(B) =82$ and supports a 30-factor gauge group
$SU(2)^{14} \times G_2^{8} \times SU(3)^{3} \times F_4^{4} \times
SO(8)$, which contains six $SU(3) \times SU(2)$ factors with jointly
charged matter.}
\label{f:typical-NHCs}
\end{figure}

This means that a typical base in 4d F-theory contains very large
non-Higgsable clusters, which contain most of the gauge factors SU(2),
SU(3) and $G_2$.  A sample of the set of non-Higgsable clusters for a
typical base encountered in one of the Monte Carlo runs is shown in
Figure~\ref{f:typical-NHCs}.  This example illustrates the complexity
of the large clusters.  This base supports a non-Higgsable gauge group
with 30 non-Abelian simple factors, with one cluster of size 16, one
cluster of size 5, one cluster of size 2, and 7 isolated gauge
factors.  The clusters of size 16 and 5 illustrate the branching and
looping possibilities of NHC's for 4d F-theory models discussed in
\cite{4d-NHC}.  The gauge factor SU(3) appears 3 times.  The
non-Higgsable gauge pair SU(3) $\times$ SU(2) appears 6 times, once as
an isolated pair and five times in the cluster of size 16.  If the
non-Abelian part of the standard model were realized on the isolated
pair, all other gauge factors not broken by fluxes would play a role
in hidden dark matter sectors; if the standard model SU(3) $\times$
SU(2) came from one of the pairs in the size 16 cluster, there could
be WIMP dark matter charged under the SU(2) and dark sector SU(2),
SU(3) or $G_2$ groups. 

\begin{figure}
\begin{center}
\includegraphics[width=10cm]{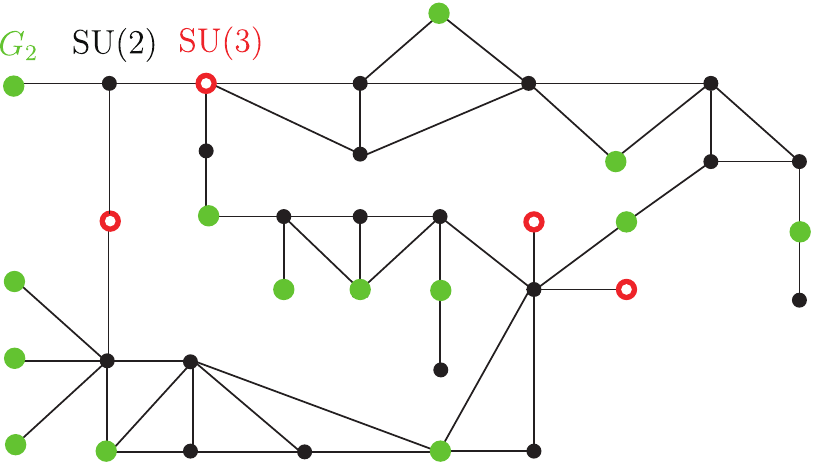}
\end{center}
\caption[x]{\footnotesize The largest non-Higgsable cluster we have encountered in our search. It contains a 37-factor gauge group
$SU(2)^{19}\times G_2^{14}\times SU(3)^4$.}
\label{f:largestNHC}
\end{figure}

Among all the $10^7$ bases encountered in the 100 Monte Carlo runs,
the largest non-Higgsable cluster was one of size 
37, with gauge group $SU(2)^{19}\times G_2^{14}\times SU(3)^4$ (see Figure~\ref{f:largestNHC}).

\subsection{Codimension two singularities and matter curves}
\label{sec:matter}

\subsubsection{Matter charged under gauge group(s)}

In F-theory the codimension-two locus of singular fibers on the base
$B$ generally corresponds to matter in the 4d supergravity theory. In
type IIB language such singularities arise at the intersection of two
sets of 7-branes. If one set of 7-branes carrys a non-Abelian gauge
group $G$, then the open strings attached between the two sets of
7-branes give rise to matter fields living in some representation $R$
of $G$. In 4d F-theory the chiral index of the charged matter in some
specific representation can be related to the G-flux in the M-theory
description of the theory \cite{Donagi-Wijnholt, bhv, bhv2, GrimmN=1, GrimmHayashi}. We
do not give any quantative description of matter curves in specific
representations here. On our 3d toric bases $B$ there are two
different types of matter curves. The first case is the toric curve
$D_i D_j$ where $D_i$ and $D_j$ possess non-Abelian gauge groups $G_1$
and $G_2$ respectively. Then generally there will be quiver matter in
the representation $(R_1,R_2)$ of $G_1$ and $G_2$.  For toric
constructions the matter is generally in the bifundamental
representation. The second case is when the vanishing locus of $\Delta$ on
a divisor $D_i$
contains a curve $C$, where $D_i$ possesses
a non-Abelian gauge group $G$,
but no other divisor carrying a non-Abelian factor passes through $C$. Then
there will be matter charged under the single gauge group $G$. This
kind of matter appears typically when the leading coefficient $\Delta_p$
in the
expansion 
\be 
\Delta=\Delta_p w^p+\Delta_{p+1}w^{p+1}+\dots 
\ee
around the divisor $D =\{w = 0\}$
contains more than one monomial. For a given divisor $D_i$, those two
different types of matter curves may simultaneously exist.

We list the proportion of gauge groups with those two types of matter
in Tables~\ref{t:singmatter}-\ref{t:anymatter}. One can see that almost
all the divisors with non-Higgsable gauge groups can have charged
matter on them.

\begin{table}
\begin{center}
\begin{tabular}{|c|c|c|c|c|}
\hline
\hline
SU(2)$_{III}$& SU(2)$_{IV}$&SU(2)&SU(3)&$G_2$\\
\hline
$93$ & $99.98$ & $97$ & $93$ & $99.6$\\ 
\hline
\hline
SO(7)&SO(8)&$F_4$&$E_6$&$E_7$\\
\hline
 $100$ 
&$70$ & $99.985$ & $94$ & $84$\\
\hline
\hline
\end{tabular}
\end{center}
\caption[x]{\footnotesize Percentage of gauge groups with matter
  charged under this single gauge  factor.}
\label{t:singmatter}
\end{table}

\begin{table}
\begin{center}
\begin{tabular}{|c|c|c|c|c|c|}
\hline
\hline
SU(2)$_{III}$& SU(2)$_{IV}$&SU(2)&SU(3)&$G_2$&$SO(7)$\\
\hline
$98.7$ & $95$ & $97$ & $86$ & $84$ & 0\\ 
\hline
\hline
\end{tabular}
\end{center}
\caption[x]{\footnotesize Percentage of gauge groups with quiver
  matter charged under this gauge factor, or equivalently, gauge groups inside some non-Higgsable cluster.}
\label{t:quivmatter}
\end{table}

\begin{table}
\begin{center}
\begin{tabular}{|c|c|c|c|c|}
\hline
\hline
SU(2)$_{III}$& SU(2)$_{IV}$&SU(2)&SU(3)&$G_2$\\
\hline
$99.987$ & $99.999$ & $99.994$ & $99.90$ & $99.95$\\ 
\hline
\hline
SO(7)&SO(8)&$F_4$&$E_6$&$E_7$\\
\hline
 $100$ 
&$70$ & $99.99$ & $94$ & $84$\\
\hline
\hline
\end{tabular}
\end{center}
\caption[x]{\footnotesize Percentage of gauge groups with any type of matter.}
\label{t:anymatter}
\end{table}

We have also counted the average number of 
each type of
possible  dark sector gauge groups adjacent to each of the SU(2)
factors in a gauge pair SU(3) $\times$ SU(2). 
Such gauge factors are generally associated with charged matter that
is also charged under the SU(2) factor.
These are listed in
Table~\ref{t:darkgaugegroup}. Similar to the distribution of gauge
pairs, the gauge groups SU(2) and $G_2$ are  dominant here.

\begin{table}
\begin{center}
\begin{tabular}{|c|c|c|}
\hline
\hline
SU(2)&SU(3)&$G_2$\\
\hline
$0.86\pm 0.23$ & $0.07\pm 0.04$ & $0.74\pm 0.19$\\
\hline
\hline
\end{tabular}
\end{center}
\caption[x]{\footnotesize Average number of each type of
``dark sector''
gauge  factor connected to each SU(2) in gauge pairs SU(3) $\times$
SU(2), with standard deviation computed among the 100 runs.} 
\label{t:darkgaugegroup}
\end{table}

\subsubsection{Codimension two singularities without gauge groups}

Besides those possibilities, there are also enhanced codimension-two
singularities without any gauge group. For example we consider the
following base $B_1$, a $\P^1$ bundle over $\F_1$, with 6 toric divisors: 
\be
v_1=(0,0,1)\ ,\ v_2=(0,0,-1)\ ,\ v_3=(0,1,0)\ ,\ v_4=(1,0,0)\ ,\ v_5=(0,-1,0)\ ,\ v_6=(-1,-1,2). 
\ee
 The set of 3d cones is:
\be
\{v_1 v_3 v_4,v_2 v_3 v_4,v_2 v_4 v_5,v_2 v_5 v_6,v_1 v_3 v_6,v_1 v_4
v_5,v_1 v_5 v_6,v_2 v_3 v_6\}. 
\ee
The orders of vanishing for $f$, $g$ and $\Delta$ on each of these
divisors are identically zero, so there are no non-Higgsable gauge
group factors supported on any toric divisors in the generic elliptic
fibration over the base $B_1$. However, on the toric curve 
$v_1 v_5$, $f$, $g$ and $\Delta$ vanish to order $(2,3,6)$. We can explicitly
write down the Weierstrass form near this curve $s=t=0$:
\be
y^2=x^3+(a_0 s^2+a_1 st^2+a_2 t^4)x+(b_1 s^3+b_2 s^2 t^2+b_3 s t^4+b_4 t^6)\label{Wform23}
\ee
The discriminant
\be
\Delta=c_1 s^6+c_2 t^{12}+\dots
\ee
has a cusp at the point $s=t=0$. In type IIB language this
configuration corresponds to many $I_1$ 7-branes intersecting on a
toric curve on $B_1$, and there is singularity enhancement on that
curve. This is a novel type of singularity, which may not be described
by the standard Kodaira ADE classification.  Systematic approaches to
resolving codimension two singularities have been described in
\cite{Katz-Vafa, mt-singularities, Esole-Yau, Lawrie-sn, Hayashi-ls,
  hlms, Esole-sy, Braun-sn}.  For the Weierstrass form
(\ref{Wform23}), however, it seems that
the singularity at $x=y=s=t=0$ cannot be
resolved using these methods. Another available technique that may be useful
in understanding these singularities is the string junction method,
which involves the (non-K\"{a}hler) deformation of the Weierstrass
model \cite{ghs3}.  It seems, however, that these represent
singularities that cannot be resolved to a total space that is a
Calabi-Yau fourfold.  The physical relevance of those codimension-two
singularities is also not clear. Naively they correspond to some
localized neutral matter.  Codimension two and higher
singularities without an apparent Calabi-Yau resolution have arisen in
several other contexts in F-theory.  Codimension 3 singularities where
$f, g$ vanish to orders of at least (4, 6) but less than (8, 12) do
not have a simple Calabi-Yau resolution but may be benign
\cite{codimension-3}.  Codimension two singularities without a CY
resolution have appeared associated with matter charged under discrete
gauge groups \cite{Braun-Morrison, mt-sections, Anderson-ggk,
  Mayrhofer-ptw, Klevers-discrete, Klevers-discrete-2}.  Codimension two 
singularities without a CY resolution are also encountered in
generalizations of the Schoen construction
\cite{Morrison-Park-Taylor}.  It seems likely that many of these
singularities are benign from the F-theory point of view, although
they may require a more sophisticated method of analysis from the
usual perspective of M-theory on a smooth Calabi-Yau; for example they
may represent cycles that are driven to vanish by their curvature, in
any supersymmetric vacuum.  Singularities of this type that cannot be
resolved to a Calabi-Yau total space were considered in
\cite{Aspinwall:1995rb}.  These codimension two singularities that
we encounter on toric bases, which do not admit a flat Calabi-Yau
resolution, may simply be a necessary feature of general 4d F-theory
models.  This kind of possibility was also discussed previously in
\cite{Klemm-lry} in a related context, where it was pointed out that
even when there is no geometric Calabi-Yau resolution of a singular
model, sensible physical features of the model can be computed for
{\it e.g.} Landau-Ginzburg models.  Thus, we proceed under the
assumption that these codimension two singularities are acceptable
features of 4d F-theory models, though their physical interpretation
remains to be fully elucidated.

In the current Monte Carlo approach we have analyzed the frequency of
occurrence of those toric cusp curves. We count the number of toric
curves $v_i v_j$ where
$\text{ord}_{D_i}(f)=\text{ord}_{D_i}(g)=\text{ord}_{D_j}(f)=\text{ord}_{D_j}(g)=0$,
$\text{ord}_{D_i D_j}(f)\geq 1$ and $\text{ord}_{D_i D_j}(g)\geq
2$. Averaging over the 100 unbounded runs, there are $2.7\pm 1.6$
toric curves of this type on each base, which implies that this
phenomenon is quite general.  Thus, a typical base in the set we have
considered has some unusual codimension two singularities that do not
appear to admit a smooth resolution to give a total space that is
Calabi-Yau, but which seem likely to be acceptable features of 4D
F-theory geometries.  

\section{Conclusions} 
\label{sec:conclusions}

We have used a Monte Carlo approach to explore a large class of
threefold bases for F-theory compactifications to four dimensions.
The bases we have considered are smooth toric threefolds that are
connected through a series of blow-up and blow-down transitions to
$\P^3$ without passing through intermediate bases with $(4, 6)$
curves.  We estimate that this set ${\cal C}$ contains on the order of
$10^{48}$ distinct threefold bases.  This is much larger than the
known and
fully enumerated set of roughly $10^4$ analogous
connected toric base surfaces that
support elliptic Calabi-Yau threefolds.  The generic elliptically
fibered Calabi-Yau fourfolds over the threefold bases in ${\cal C}$
give roughly $10^{48}$ elliptically fibered Calabi-Yau fourfolds.
While some fourfolds may admit multiple distinct elliptic fibrations, this number still should act as a reasonable
lower bound for the number of possible distinct elliptic Calabi-Yau
fourfolds.  
Modifying the approach used here slightly, it is straightforward to
systematically construct all the toric threefold bases in the
connected set up to any given value of $h^{1, 1}(B)$.  Such an analysis
could be implemented and would be bounded only by computational
resources; for example, computing the first $10^9$ bases would reach
to roughly $h^{1, 1}(B) \cong 10$.
We have considered the gauge groups and matter that are
supported by geometric non-Higgsable clusters over the bases explored.
A typical base has a non-Higgsable gauge group with roughly 30
factors, dominated by SU(2) and $G_2$, with some SU(3) and other
factors arising.  Roughly 10\% of connected group factor
pairs are SU(3)
$\times$ SU(2) pairs.

The set that we have explored here represents an enormous family of
elliptic Calabi-Yau fourfolds.  For elliptic Calabi-Yau threefolds, it
is known that the number of distinct topological types is finite
\cite{Gross}, and that all are connected through extremal transitions,
through the minimal model theory for the base surfaces \cite{Grassi}.
This makes possible in principle a complete and systematic
classification of all elliptic Calabi-Yau threefolds, for which the
numbers involved do not seem prohibitive and towards which substantial
progress has been made \cite{mt-toric, Hodge, Martini-WT, Johnson-WT,
  Wang-WT}.  For fourfolds, however, beyond the apparently prohibitive
size of the number of distinct topologies involved, there are a number
of further theoretical steps needed to get a systematic handle on the
set of possibilities.  We mention here briefly some things that were
not done in this work that would represent further progress in this
direction.  First, there is no proof of finiteness for elliptic
Calabi-Yau fourfolds, and the analogue of the minimal surfaces for
threefold bases has not been worked out systematically.  While the
observation that the set we have explored here connects together all
the known toric threefold bases with small $h^{1, 1}(B)$ except those
with $E_8$ factors that cannot be reached except through intermediate
threefolds with $(4, 6)$ curves, suggesting that all toric threefold
bases may be connected through extremal transitions, this has not been
proven even with the restriction to toric structure.  
It would be nice at least
to generalize the analysis done here to include
almost-toric bases that have $E_8$ factors and $(4, 6)$ curves, as has
been done in the case of base surfaces, since such bases seem to play
an important role for threefolds and fourfolds at large Hodge numbers.
Over each given
threefold base, in general a variety of elliptic Calabi-Yau fourfolds
can be constructed by tuning different codimension one and two
singularities, corresponding to Higgsable gauge groups and matter in
the F-theory picture.  For many threefold bases this can give rise to
a vast array of distinct elliptic Calabi-Yau fourfolds.  A systematic
analysis of such tuning would help to indicate how much larger the
complete set of elliptic fourfolds might be than the set of base
threefolds considered here.  A systematic approach to tuning in the case
of elliptic CY threefolds was described, for example, in
\cite{Johnson-WT}.  Another question is the extent to which toric
threefolds are a representative sample of the complete set of
threefold bases that support elliptic Calabi-Yau fourfolds.  For
elliptic CY threefolds, a systematic analysis of non-toric bases
\cite{Wang-WT} shows that, at least at large $h^{2, 1}$, toric base
surfaces form a good representative sample of the full set of
non-toric bases, but a similar analysis for threefold bases would be
substantially more complex as it would involve blowing up curves in
addition to points.

Finally, a few words regarding the physical relevance of the
distribution we have sampled here.  The Monte Carlo we have carried
out samples each distinct threefold base $B$ with a weight
proportional to its number of allowed neighbors, which we have used to
compute statistical averages based on an equal weighting of each toric
threefold base.  There is no physical reason why this is a correct
weighting, this is simply a mathematical formulation of a simple
weighting factor that allows us to study typical features of the
ensemble based on an equal weighting of base threefolds.  A proper
weighting of bases for physics is not fully understood and would
depend on the detailed global dynamics of string theory. The effects
of G-flux and world-volume brane dynamics would also need to be
included to systematically analyze the set of possibilities from a
physics perspective.  The most plausible approach advanced so far for
understanding physical vacuum distributions in F-theory is the
statistical approach to flux vacua developed by Ashok, Douglas and
Denef \cite{Ashok-Douglas, Denef-Douglas} (see \cite{Braun-Watari,
Watari} for a
recent application in the F-theory context, and
\cite{Douglas-Kachru, Denef-F-theory} for pedagogical reviews).  
We leave a consideration of the effects of fluxes on the
distribution of vacua in the context developed here for future work.

{\bf Acknowledgements}: We would like to thank Lara Anderson, 
Andreas Braun,
James
Gray, Thomas Grimm,
Jim Halverson, and David Morrison for helpful
discussions.  We would also like to thank Jan Balewski for his
assistance and the use of the ``reuse'' computer cluster.
This research was supported by the DOE under contract 
\#DE-SC00012567.

\end{document}